# Magnetospheric Multiscale (MMS) observations of foreshock transients at their very early stage

Terry Z. Liu[1,2], Xin An[3], Hui Zhang[2], and Drew Turner[4]

[1]Cooperative Programs for the Advancement of Earth System Science, University Corporation for Atmospheric Research, Boulder, CO, USA. [2]Geophysical Institute, University of Alaska, Fairbanks, Fairbanks, AK, USA. [3]Department of Atmospheric and Oceanic Sciences, University of California, Los Angeles, Los Angeles, CA, USA. [4]Johns Hopkins University Applied Physics Laboratory, Laurel, Maryland, USA.

**Abstract**

Foreshock transients are ion kinetic structures in the ion foreshock. Due to their dynamic pressure perturbations, they can disturb the bow shock and magnetosphere-ionosphere system. They can also accelerate particles contributing to shock acceleration. However, it is still unclear how exactly they form. Recent particle-in-cell simulations point out the important role of electric field and Hall current in the formation process. To further examine this, we use data from the Magnetospheric Multiscale (MMS) mission to apply case studies on two small (1000-2000 km) foreshock transient events that just started to form. In event 1 where MMS were in a tetrahedral formation, we show that the current density configuration, which determined the magnetic field profile, was mainly driven by Hall currents generated by demagnetized foreshock ions. The resulting time variation of the magnetic field induced electric field that drove cold plasma moving outward with magnetic field lines. In event 2 where MMS were in a string-of-pearls formation, we analyze the evolution of field and plasma parameters. We show that the magnetic flux and mass flux were transported outward from the core resulting in the steepening of the boundary. The steepened boundary, which trapped more foreshock ions and caused stronger demagnetization of foreshock ions, nonlinearly further enhanced the Hall current. Based on our observations, we propose a physical formation process that the positive feedback of foreshock ions on the varying magnetic field caused by the foreshock ion Hall current enables an "instability" and the growth of the structure.

## 1. Introduction

Upstream of Earth's quasi-parallel bow shock, the ion foreshock is characterized by backstreaming ions that have been reflected from the shock (e.g., Eastwood et al., 2005; Wilson, 2016). In the ion foreshock, many foreshock transients have been observed and simulated, such as hot flow anomalies (HFAs) (e.g., Schwartz et al., 1985, 2018; Thomsen et al., 1986, 1988; Thomas et al., 1991; Lin, 1997; Omidi and Sibeck, 2007; Zhang et al., 2010), spontaneous hot flow anomalies (Omidi et al., 2013; Zhang et al., 2013), foreshock bubbles (FBs) (e.g., Omidi et al., 2010; Turner et al., 2013, 2020; Liu et al., 2015), foreshock cavities (e.g., Lin, 2003; Sibeck et al., 2002), foreshock cavitons (e.g., Blanco-Cano et al., 2011) and short large-amplitude magnetic structures (SLAMS) (e.g., Schwartz et al., 1992; Wilson, 2016). HFAs, SHFAs, and FBs are three of the most significant types of foreshock transients due to their large sizes (e.g., several $R_E$ for HFAs/SHFAs and even larger for FBs), strong perturbations, and plasma heating. They are characterized by a hot, tenuous core associated with plasma deflection bounded by compressional boundaries or shocks on one or both sides. Due to large variation in plasma density and velocity, the dynamic pressure is distinct from the surrounding solar wind and foreshock plasma. As a result, when these foreshock transients encounter the bow shock, the bow shock surface can locally move back and forth. Such perturbation can propagate to the magnetopause causing magnetospheric and ionospheric disturbances (e.g., Hartinger et al., 2013; Archer et al., 2014, 2015; Sibeck et al., 1999; Zhao et al., 2017; Wang et al., 2018).

Recent observations showed that foreshock transients can also accelerate particles (e.g., (Kis et al., 2013; Wilson et al., 2013, 2016; Liu et al., 2017a). When HFAs or FBs expand supermagnetosonically, a shock can form. Such shocks can accelerate solar wind particles through shock drift acceleration (Liu et al., 2016a). When the boundary of HFAs/FBs convects towards the bow shock, ions and electrons can bounce between the two regions of strong compressed magnetic fields, resulting in Fermi acceleration (Liu et al., 2017b; Liu et al., 2018; Turner et al., 2018). As magnetic flux is transported towards the boundary during the expansion of an FB, electrons can be accelerated through betatron acceleration (Liu et al., 2019). Recently, magnetic reconnection was observed to heat electrons inside HFAs/FBs and SLAMS (Liu et al., 2020; Wang et al., 2020). Shock acceleration is one important acceleration mechanism but is still not fully understood (see review by Treumann, 2009). For example, the particle acceleration efficiency is underestimated and the source of energetic particles that can participate in the acceleration process is unclear. Foreshock transients, which are often present upstream of supercritical shocks, could potentially increase particle acceleration efficiency and provide a particle source (e.g., Turner et al., 2018).

However, how HFAs, SHFAs, and FBs form is still not fully understood. In the simulations by Omidi et al. (2013), SHFAs form from foreshock cavitons (the nonlinear evolution of ULF waves), but the mechanism is unknown. For HFAs and FBs, thermal pressure enhancement by foreshock ions is considered to drive their formation and expansion, which, however, is insufficient. Based on simulations (e.g., Burgess, 1989; Thomsen et al, 1989; Lin, 2002; Omidi and Sibeck, 2007; Omidi et al., 2010), HFAs and FBs form when foreshock ions are trapped by a solar wind discontinuity. Around certain magnetic field configurations of solar wind discontinuities (e.g., Archer et al., 2015; Liu et al., 2015), the gyro-kinetic motion of foreshock ions can result in their concentration and thermalization, resulting in a thermal pressure increase. The increased thermal pressure can push ambient cold plasma outward forming a low-density core surrounded by compressional boundaries that, depending on the expansion speed into the surrounding plasma, can form into fast mode shocks. However, as the gyroradius (1000s of km) and gyroperiod (10-20s) of foreshock ions are larger than or comparable to the spatial scale and time scale of field variation around HFAs/FBs, respectively, the concept of thermal pressure of foreshock ions is invalid. Therefore, the kinetic effects of foreshock ions must be considered.

Recent particle-in-cell (PIC) simulations provide a physical model (An et al., 2020 ApJ accepted). When foreshock ions encounter a discontinuity, foreshock ions are demagnetized whereas electrons are magnetized resulting in a Hall current which shapes the magnetic profile of a foreshock transient. The associated electric field transfers energy from foreshock ions to cold plasma and the field. To confirm and further investigate these, it is important to observe those foreshock transients that just start to form. During their very early stage, they must be very small (e.g., less than or comparable to one foreshock ion gyroradius) and may evolve very fast. Only recently have we the sufficient time resolution of particle measurements to resolve them using NASA's Magnetospheric Multiscale (MMS).

Using MMS, we study two very small foreshock transients (around 5s in duration; 1000-2000 km in size). We do not distinguish SHFAs, HFAs, or FBs in this study because their driver discontinuities are difficult to identify when embedded in the ULF waves and their distinctive characteristics (e.g., size and upstream shock) are not available when they just form. At the end of this paper, however, we discuss possible differences in their formation process. In event 1

(Section 3.1), with MMS in a tetrahedral formation, we analyze how foreshock ions contributed to the current density configuration that determined the magnetic field geometry of the event. In event 2 (Section 3.2) with MMS in a string-of-pearls formation, we analyze how plasma and field parameters evolved. In Section 4, we summarize our results and propose a formation mechanism.

## 2. Data and Methods

We used data from NASA's MMS mission (Burch et al., 2016). We analyzed plasma data from the Fast Plasma Investigation instrument (FPI; Pollock et al., 2016), DC magnetic field data from the fluxgate magnetometer (Russell et al., 2016), magnetic field wave data from the search coil magnetometer (Le Contel et al., 2016), and electric field data from axial and spin-plane double-probe electric-field sensors (Ergun et al., 2016; Lindqvist et al., 2016).

During dayside seasons with apogees from 12 $R_E$ to 25 $R_E$, MMS observed many SHFAs, HFAs, and FBs. We searched for events that have very short duration (a few seconds) observed in burst mode with very high resolution (30 ms for electrons and 150 ms for ions). Here we present case studies on two representative events with spatial scale ~1000-2000 km (comparable to around one foreshock ion gyroradius or 10-25 ion inertial length). In event 1, the four identical MMS spacecraft were in a tetrahedron formation with very small separation of ~20-30 km. Such a formation provides the availability of the four-spacecraft timing method (Schwartz, 1998) and the curlometer method (Robert et al., 1998). In event 2, MMS spacecraft were in a string-of-pearls formation with separation from 200 km to 400 km. Such a formation can capture the fast evolution of event 2 within 1s.

## 3. Results

### 3.1. Event 1: Current and Field Configuration

In Figure 1, MMS observed a foreshock transient at the flank of the bow shock ([6.0, -14.7, 4.1] $R_E$ in GSE). It had the common characteristics of typical SHFAs/HFAs/FBs except for the very small size (1000-2000 km along the GSE-X direction comparable to one foreshock ion gyroradius). The transient had a core with low field strength (Figure 1a), low density (Figure 1b), and plasma deflection (Figure 1c) associated with electron heating (Figure 1f) (ion temperature is not shown as it is inaccurate in the foreshock). Upstream of the core, there was a compressional boundary with enhanced field strength and density (vertical dotted line). Using the four-spacecraft timing method (Schwartz, 1998), the boundary normal was [0.95, -0.29, -0.03] in GSE and the normal speed was -303 km/s nearly the same as the local ion bulk velocity along the normal direction in the spacecraft frame. At the boundary (vertical dotted line), the electron perpendicular temperature shows an increase profile similar to that of the field strength, suggesting the betatron acceleration consistent with Liu et al. (2019). The interplanetary magnetic field (IMF) variation across the event was not significant. Before the event, the IMF was dominated by $B_y$ in GSE. After the event, $B_y$ became slightly weaker and other two components became slightly stronger. If we assume there was a tangential discontinuity (TD), its magnetic shear angle was only ~25° and its normal was [0.56, -0.17, 0.81] in GSE calculated from the cross product method (Schwartz et al., 1998) using the time interval between the vertical dashed lines before and after the event. (Unfortunately, no ARTEMIS, Cluster, or Geotail was available in the upstream solar wind and it is difficult to identify a discontinuity with a small shear angle at the L1 point.) This event could be either a foreshock caviton-driven SHFA or a solar wind discontinuity-driven HFA/FB. The geometry of the event is sketched in Figure 2a, and the magnetic field configuration is sketched in Figure 2b, which shows curved field lines corresponding to two $B_z$ reversals in Figure 1a.

As the magnetic field structure of the foreshock transient was determined by the current density configuration, to understand how the foreshock transient formed it is important to examine the current density and how motions of ions and electrons contributed to it. Figures 3b and 3c show the current density calculated from the curlometer method (Robert et al., 1998) and from plasma data, respectively. They are qualitatively consistent except the upstream region (gray shaded region). We see that there was overall negative $J_y$ in the core with two peaks at two edges of the core (2$^{nd}$ and 4$^{th}$ vertical dotted line at 16:30:01 UT and 16:30:03.5 UT, respectively; note that the fluctuation at the 2$^{nd}$ vertical dotted line in Figure 3b was whistler waves). Such negative $J_y$ (purple out-of-plane symbol in Figure 2) was likely the reason of $B_z$ reversal from negative to positive in the core. There were also other currents. At the leading edge (downstream) of the core (16:30:01 UT), there were finite $J_x$ and $J_z$ likely responsible for $B_y$ depletion in the core. At the trailing (upstream) edge of the compressional boundary (16:30:04 UT), there was positive $J_y$ peak, which caused $B_z$ reversal from positive to negative. The positive $J_y$ peak could close a current loop with the negative $J_y$ peak at 4$^{th}$ vertical dotted line and together enhanced positive $B_z$ at the compressional boundary. Inside the compressional boundary (16:30:03.5-16:30:04 UT; 5$^{th}$ vertical dotted line), there was negative $J_z$ which was likely responsible for the reversal of $B_x$ and $B_y$.

Next, we determine what caused such current density configuration by examining the velocity of foreshock ions, solar wind ions, and electrons inside the event (16:30:01-16:30:04 UT). Figures 3d and 3e show the total ion bulk velocity and electron bulk velocity, respectively. They were similar overall, but in the core ion $Vi_y$ was negative whereas electron $Ve_y$ was around zero or positive, resulting in the negative $J_y$. To examine the reason of this velocity difference, we calculated the velocity of solar wind ions and foreshock ions separately by confining the energy and direction from ion distributions (Figures 3f and 3g). Unlike total ion bulk velocity, the solar wind ion velocity in the core was very similar to the electron bulk velocity. To see it more clearly, Figure 3h shows their comparison in the perpendicular velocity. We see that electron and solar wind ion perpendicular velocities almost overlap (solid and dashed lines; except in the upstream region). Total ion perpendicular velocity (dotted line in Figure 3h), on the other hand, clearly shows smaller GSE-Y component in the core compared to electrons. The reason is that foreshock ions were moving mainly in the negative GSE-Y direction (Figure 3g). Therefore, the differences between total ion bulk velocity and electron bulk velocity and thus the current density were mainly due to the motion of foreshock ions.

To further examine the motion of foreshock ions, we plot ion distributions shown in Figure 4. In the background foreshock (1st vertical dotted line in Figure 3), Figure 4a shows that in the GSE-XY slice, foreshock ions were mainly moving in the negative GSE-X and GSE-Y direction. Figure 4c shows that in the BV slice (horizontal axis is along the magnetic field and the plane contains the electron bulk velocity vector), the center of the foreshock ion distribution shows E×B drift same as the solar wind ions and a parallel component opposite to the solar wind ions. In the perpendicular slice (Figure 4d; cut through the vertical dotted line in Figure 4c), foreshock ions show a complete gyration. Therefore, the background foreshock ions were moving along the field lines with large thermal speed. Because the IMF was dominated by negative $B_y$, the foreshock ion bulk velocity therefore was dominated by a negative GSE-Y component and had a GSE-X component similar as the solar wind ions (E×B drift) consistent with Figure 3g. Such foreshock ion motion cannot cause strong current in the background foreshock because electrons can move freely along the field lines (Figure 3b).

In the core (2$^{nd}$ and 3$^{rd}$ vertical dotted lines in Figure 3), foreshock ions were still mainly moving in the negative GSE-X and GSE-Y direction (Figures 3g, 4e and 4i). This is because ion gyroperiod was very long (~10-20s) compared to the time scale of field variation (e.g., within 1s in event 2), foreshock ions cannot immediately change their velocity, i.e., foreshock ions were demagnetized. Furthermore, because the magnetic field direction varied from $B_y$ dominant to $B_z$ dominant (Figure 3a), the foreshock ion bulk velocity, which was initially field-aligned, projected to the perpendicular direction. As a result, in the perpendicular slices (Figures 4h and 4l), foreshock ions changed from a complete gyration to partial gyration in the direction opposite to the convection electric field (i.e., in the negative GSE-Y direction, see Figure 3k). Such perpendicular velocity caused the negative $J_y$ (Figures 3b and 3c) as electrons were nearly always magnetized. In other words, it was the Hall current driven by the demagnetized foreshock ions.

Because the gyroradius of foreshock ions was comparable to the spatial scale of the event, the initial gyration of foreshock ions can also contribute to the Hall current in the GSE-XZ plane. In the BV slices (Figure 4g and 4k), we see that foreshock ion velocity had a sunward component diffuse in the field-aligned direction, because some of the (sunward) gyrovelocity of foreshock ions projected to the new field-aligned direction (orange arrow in Figure 2b). We also see that there was less earthward gyration in Figures 4g and 4k compared to Figure 4c, which can also be seen in the perpendicular slices by comparing with the gyrocenter (red dots in Figures 4h and 4l), in the XZ slices (Figures 4f and 4j), and in Figure 3g (weaker GSE-X component). The possible reason is that as the event had a compressional boundary on its upstream side and no compressional boundary on its downstream side, fewer foreshock ions from the upstream side can enter the core and contribute earthward gyration than those from the downstream side that contributed sunward gyration. This preference in gyrophase caused a velocity difference in the GSE-X direction between foreshock ions and electrons (Figure 3h) and thus small Hall current $J_x$ (around 2$^{nd}$ vertical dotted line in Figure 3b). (There was also small $J_z$ right before $J_x$ (Figure 3b), probably because there were more foreshock ions entering the core than leaving the core from the downstream side.) Additionally, at the 3$^{rd}$ vertical dotted line, the calculated foreshock ion density dropped from 0.5 to 0.35 cm$^{-3}$ (Figure 3j), possibly due to the same reason as $V_x$ variation (foreshock ions from the upstream side were obstructed by the compressional boundary).

Next, we examine the $J_y$ peak at the edge of the compressional boundary (4$^{th}$ vertical dotted line in Figure 3). We see that foreshock ion velocity was still in the negative GSE-X and Y direction except that more earthward gyration appeared (Figures 4m-p) likely because more foreshock ions from the upstream side can reach here. Note that the reversal of the foreshock ion distribution in the perpendicular slice (Figure 4p) was simply caused by the reversal of $B_z$ (and thus **B**×**V**). This $J_y$ peak was due to at least two reasons. One reason is the enhancement of foreshock ion density (Figure 3j) which was likely because the field strength enhancement can help trap foreshock ions by preventing them from gyrating away. This is the same reason that caused foreshock ion density depletion in the core (less foreshock ions from the upstream side can gyrate further downstream into the core). The other reason is the electron E×B drift in the positive GSE-Y direction (Figures 3e and 3h). At the edge of the boundary with strong magnetic field gradient, there was an enhancement in electric field $E_x$ (Figure 3k). As the boundary normal was mainly in the sunward direction, the enhanced negative $E_x$ was pointing from the boundary into the core. Our calculation shows that $-E_xB_z$ component dominates the (**E**×**B**)$_y$. The possible process is that as ions (especially foreshock ions) from the core can penetrate further into the field strength enhancement region than electrons due to their different gyroradius, static electric field arose pointing from the boundary into the core (sketched in Figure 2b). Such electric field can drive electrons E×B drift in the GSE-Y direction but had nearly no effects on foreshock ions resulting in the Hall current along the boundary surface (this electric field component was corresponding to the Hall electric field). This process is similar to that at shock surfaces (see review in Treumann, 2009) and can also explain the positive $J_y$ peak at the upstream edge of the compressional boundary (Figure 3b; although we do not have accurate plasma data to confirm this current). Such a scenario is consistent with recent PIC simulations by An et al. (2020 ApJ accepted).

As for negative $J_z$ in the middle of the boundary ($5^{th}$ vertical dotted line in Figure 3), although electrons show clear enhancement in $Ve_z$ and ions do not (Figures 3d-g), their perpendicular velocities match very well in the GSE-Z direction (Figure 3h). Therefore, their velocity difference in the GSE-Z direction was dominant in the field-aligned direction. The reason is likely that as magnetic field and electric field evolved rapidly at the compressional boundary (see event 2), electrons can immediately respond and change their drift velocity (electron bulk velocity at the compressional boundary was almost the same as the perpendicular velocity), but ions cannot. The varied magnetic field projected some ion velocity to the parallel direction resulting in field-aligned current. The effect of this field-aligned current was to twist the magnetic field lines and make them look like a flux rope but cut in half (see sketch in Figure 2b). (As the boundary continued to steepen, the open part of the "half flux rope" might have a chance to close, i.e., magnetic reconnection might occur and form a real flux rope. This is a possible explanation of the observations of a small-scale flux rope at an HFA's compressional boundary by Bai et al. (2020))

Here we summarize a possible formation process based on our observations: Initially, when foreshock ions encountered a B-field discontinuity (e.g., a solar wind discontinuity or a steepened ULF wave/foreshock caviton), they cannot immediately change their bulk velocity and cannot complete a gyration. In other words, as the time scale of field variation was shorter than ion gyroperiod and the discontinuity thickness/field variation spatial scale was less than or comparable to a foreshock ion gyroradius, foreshock ions were demagnetized whereas electrons were nearly always magnetized resulting in the establishment of a Hall current (purple vectors in Figure 2b). Such Hall current varied the magnetic field around the discontinuity. We propose that if such field variation favors stronger Hall current, the stronger Hall current can in return further vary the magnetic field. This completes a feedback loop resulting in a kind of instability and thus the nonlinear growth of a foreshock transient (SHFA, HFA, and FB). We will examine this process in event 2.

The field variation (or field line motion) during this process induced an additional electric field (e.g., enhanced $E_y$ shown in Figure 3k). The effect of this electric field was to self-consistently drive frozen-in cold plasma moving outward together with the magnetic field lines (red arrows in Figure 2b). Such outward motion was corresponding to the expansion of the foreshock transient consistent with PIC simulations by An et al. (2020 ApJ accepted). As sketched in Figure 2b (in the solar wind rest frame), as MMS transited (gray dashed arrow), solar wind ion and electron perpendicular velocity (Figure 3h) and E×B drift velocity (Figure 3j) first had a large negative GSE-Z component corresponding to the deformation towards the downstream side (red arrows; also see Figure S1 in the supporting information showing the electron bulk velocity in the solar wind rest frame). Later, the field line direction rotated, and the cold plasma motion became mainly sunward (in the solar wind rest frame) resulting in the compression at the upstream boundary.

Finally, we discuss the energy budget as both the cold plasma outward motion and magnetic field deformation require energy input. As the process started from the Hall current driven by the demagnetized foreshock ions, foreshock ions must be the energy source. This can be seen from the enhanced +$E_y$ around the leading edge of the event and foreshock ion -$V_y$, i.e., the partial gyration of foreshock ions was against the induced electric field in the core. At the compressional boundary, however, the enhanced $E_y$ was negative meaning that foreshock ions were gaining energy. This is simply because the observation was in the spacecraft frame where foreshock ions were pushed back by the earthward moving compressional boundary (Liu et al., 2018). To consider the energy conversion between the foreshock ions and the field, we need to exclude the effect of background convection electric field. In the solar wind rest frame (see Figure S1 showing $\boldsymbol{E} + \boldsymbol{V}_{sw} \times \boldsymbol{B}$), $E_y$ was positive (and cold plasma $V_x$ became sunward), and foreshock ions were losing energy by pushing the compressional boundary sunward consistent with PIC simulations by An et al. (2020 ApJ accepted).

### 3.2. Event 2: Evolution

Figure 5 shows observations of event 2 by four MMS spacecraft. In this event, MMS were in a string-of-pearls formation (see geometry in Figure 6). MMS2 was the first spacecraft that observed the event (Figure 5.1). Similar to event 1, it was a very small (~2000 km along the GSE-X direction) foreshock transient with a hot, tenuous core associated with plasma deflection bounded by a compressional boundary on the upstream side. Inside the core, foreshock ion energy flux shows energy dispersion (Figure 5.1d). Because the foreshock magnetic field was fluctuating significantly, we cannot determine the driver discontinuity in this event. MMS1, which was ~200 km further downstream than MMS2, observed the event around 0.4s later (Figure 5.2). In MMS1 observations, the compressional boundary steepened, and a wave train appeared right upstream of the compressional boundary (Figure 5.2a) which was very likely a whistler precursor often observed upstream of shocks and SLAMS (e.g., Wilson, 2016). Inside the core, the foreshock ion energy flux enhanced (from yellow to red), whereas the solar wind ion energy flux depleted (less red), and the solar wind ion energy decreased a little (Figure 5.2d). The electron temperature in the core further increased (Figure 5.2f). When MMS4 (~100 km downstream than MMS1) observed the event (Figure 5.3) around 0.2s later, the compressional boundary further steepened, and the amplitude of the wave train increased. The foreshock ion energy flux further enhanced, while the solar wind ion energy flux further depleted, and the solar wind energy further decreased (Figure 5.4d). (Unfortunately, MMS4 FPI electron measurement became unavailable since mid-2018.) Using minimum variance analysis (Sonnerup & Scheible, 1998), MMS2, MMS1, and MMS4 observed similar compressional boundary normal of [0.81, -0.55, 0.19], [0.80, -0.36, 0.47], and [0.83, -0.27, 0.47] in GSE, respectively (minimum-to-intermediate eigenvalue ratio less than 0.1). Using the coplanarity method (Schwartz, 1998), we also obtained similar normal of [0.77, -0.23, 0.58], [0.75, -0.21, 0.63], and [0.74, -0.18, 0.65] in GSE, respectively. Such normal direction was roughly along the spacecraft train (less than ~25°; Figure 6). Therefore, the observation differences among three spacecraft were mainly due to the evolution rather than the spatial difference. MMS3, which was ~400 km further downstream than MMS4, observed the magnetosheath part of the event.

To see the evolution clearer, we time shifted MMS1 and MMS4 observations by -0.4s and -0.6s, respectively, to match the downstream boundary observed by MMS2 (see their superposed plots in Figure 7). By comparing their magnetic field strength and electron density (ion density had large uncertainties) shown in Figures 7a and 7c, we see there was clear magnetic flux and mass flux transport from the core towards the upstream boundary, consistent with the outward motion of plasma along with field lines discussed in event 1. Such sunward magnetic flux transport can result in betatron acceleration, which increased the electron perpendicular temperature at the inner edge of the compressional boundary (Figure 5.2f). As a result, Figure 7d shows enhanced perpendicular anisotropy at the compressional boundary corresponding to the field strength increase, consistent with Liu et al. (2019). Figures 7e and 7f show that the ion and electron deflection became more and more significant.

Figure 8 shows the ion distribution evolution from MMS2 to MMS4 (the time differences between them are roughly the same as the time shift in Figure 7; corresponding to vertical dotted lines in Figure 5). In the background foreshock, sunward foreshock ions can be seen (Figures 8a-f), as the IMF had strong radial component unlike event 1. Inside the core (Figures 8g-l), three MMS spacecraft observed earthward foreshock ions with speed faster than the sunward component. Further into the core (Figures 8m-r), the earthward foreshock ion energy became smaller corresponding to the energy dispersion observed in Figure 5d. This scenario is consistent with previous THEMIS observations (Liu et al., 2018) that sunward foreshock ions were reflected by the earthward moving compressional boundary and gained speed in the spacecraft frame (in the solar wind rest frame, foreshock ions lost energy as discussed in event 1; also see Figure S2 in the supporting information showing the ion distribution in the local flow rest frame where the foreshock ion energy does not change). The energy dispersion was due to the time-of-flight effect (faster ions reached the spacecraft earlier). Comparing distributions measured by the three spacecraft in the core, we see that the solar wind phase space density becomes lower and lower from MMS2 to MMS4, consistent with Figures 5.1-5.3d. The calculated solar wind ion density decreased from MMS2 to MMS1 in Figure 9h. The decrease in solar wind ion density was due to the sunward mass flux transport from the core. Figure 9h also shows increase in the foreshock ion density consistent with Figures 5.1d and 5.2d, likely because steepened compressional boundary can reflect and trap more foreshock ions in the core. If we compare the calculated solar wind velocity measured by MMS2, MMS1, and MMS4 (Figures 8m-r): [-283.2, -58.9, -28.5] km/s, [-232.8, -100.3, -25.0] km/s, and [-224.1, -118.5, -25.0] km/s, respectively, we see that MMS1 observed slower and deflected solar wind ions than MMS2 by ~50 km/s in both GSE-X and Y directions and MMS4 observed further deceleration/deflection by another 20 km/s (also see vertical dotted lines which indicate the solar wind speed observed by MMS2 in Figure 8), consistent with Figures 5.1-5.3d (solar wind ion energy decreased). Therefore, the stronger ion bulk velocity deflection observed by MMS1 and MMS4 than that by MMS2 (Figure 7e) was due to increasing foreshock ion density, decreasing solar wind ion density, and deceleration/deflection of the solar wind ions in the spacecraft frame (acceleration in the solar wind rest frame).

Next, we examine the evolution of the current density configuration. Limited by the uncertainty of the velocity measurement, only current density in the GSE-X direction was high enough to be seen in Figure 9.1b. This can also be seen from the ion (Figure 9.1d) and electron bulk velocity (Figure 9.1e). Their perpendicular velocities, on the other hand, only show small differences in the GSE-X direction (Figure 9.1f), indicating that the positive $J_x$ was mainly field-aligned. One possibility is that as the background IMF had strong radial component, the background solar wind ions and electrons had large field aligned speed. As the field direction varied in the core, electrons could maintain the field-aligned speed whereas ions projected the field-aligned speed to the perpendicular direction. Meanwhile, foreshock ions also contributed sunward velocity. At MMS1 (Figure 9.2), a bipolar $J_y$ signature on two edges of the compressional boundary similar to event 1 enhanced to be seen (Figure 9.2a), responsible for the steepening of the compressional boundary (see sketch in Figure 6). Like event 1, the enhancement of positive $J_y$ was caused by at least two processes. One process is the density enhancement of foreshock ions (Figure 9h) that were moving in the positive GSE-Y direction (Figure 8). Another process is that electrons can respond to the field evolution (Figures 9a and 9c) much faster than ions. As a result, electron $Ve_y$ varied more significantly than ion $Vi_y$ (Figures 9d and 9e). A possible process could be that the steepened compressional boundary can cause foreshock ions more demagnetized and stronger static electric field pointing from the boundary to the core (difficult to see it from Figure 9c as the convection electric field was dominant). The enhanced earthward static electric field can drive stronger electron E×B drift along the boundary surface but cause no effects on foreshock ions, contributing to the stronger Hall current. This process is similar to the shock steepening process which can also explain the enhancement of negative $J_y$ at the trailing edge of the compressional boundary.

Here we summarize our results. The Hall current between the demagnetized foreshock ions and magnetized electrons deformed the magnetic field configuration, which transported magnetic flux from the core to steepen the compressional boundary (Figure 7a). The enhancing field strength at the compressional boundary reflected and trapped more foreshock ions (Figures 5d, 8 and 9h), and the sharper field variation caused foreshock ions more demagnetized and stronger static electric field, resulting in a larger velocity difference between ions and electrons (i.e., a feedback loop; Figures 9d and 9e). Increases in both foreshock ion density and velocity difference can cause a stronger Hall current (Figure 9b). The stronger Hall current can further steepen the compressional boundary (Figure 7a), which can in return cause even stronger Hall current. This is the same nonlinear feedback growth process as was already outlined for event 1.

The variation in the magnetic field induced convection electric field drove cold plasma outward and caused sunward mass flux transport (Figure 7c). The outward moving speed must increase from 0 to a certain value in the solar wind rest frame meaning that there should be acceleration (or deceleration in the spacecraft frame) as shown in Figures 5d and 8. The physical process of this acceleration could be: As it was a kind of instability, the growth in the Hall current was nonlinear ($\partial^2 \boldsymbol{j}/\partial t^2 \neq 0$) resulting in faster magnetic field variation ($\partial^2 \boldsymbol{B}/\partial t^2 \neq 0$) and consequently the increasing electric field (Figure 9c; $\frac{\partial}{\partial t}(\nabla \times \boldsymbol{E}) \neq 0$). The increasing electric field can be responsible for the acceleration of the frozen-in cold plasma, e.g., through a process similar to the acceleration of pickup ions, self-consistently corresponding to the faster magnetic field line motion.

## 4. Conclusions and Discussion

Using MMS, we analyzed two very small foreshock transients in their earliest stages of development to understand how they formed. We used a tetrahedron formation to study the current density configuration inside one foreshock transient to show how motions of foreshock ions and electrons contributed to it. Then we used a string-of-pearls formation to study the temporal evolution of plasma and field parameter in another foreshock transient. Based on our observational results, we summarized a formation model as follows: When suprathermal foreshock ions encounter a B-field discontinuity (e.g., a solar wind discontinuity for HFAs/FBs or a steepened ULF wave/foreshock caviton for SHFAs), they cannot immediately change their bulk velocity (the ion gyroperiod is typically much longer than the time scale of early foreshock transient formation as shown in event 2) and cannot complete a gyration (the foreshock ion gyroradius is larger than or comparable to the spatial scale of field variation and the thickness of the B-field discontinuity), so the foreshock ions become demagnetized. As electrons are nearly always magnetized, a Hall current is established, which varies the magnetic field profile around the discontinuity. If such magnetic field variation then further enhances the Hall current (Figure 9b), e.g., by trapping more foreshock ions (Figure 9h) and causing a larger velocity difference between ions and electrons (Figures 9d and 9e), the enhanced Hall current can in return further steepen the magnetic field profile. This feedback loop forms a kind of nonlinear instability (see a simple derivation in Appendix A) that enables the growth and development of the foreshock transients. During the growth of the magnetic field profile, an induced electric field is established, which self-consistently drives frozen-in cold plasma to move outward (e.g., Figure 2b) together with the field lines (mass and magnetic flux outward transport in Figure 7a). The outward moving speed must increase from 0 meaning acceleration in the solar wind rest frame (Figures 5d and 8) driven by the enhancing electric field (Figure 9c) since the growth of the Hall current, and thus the time variation of magnetic field, are nonlinear (instability). The energy source is the foreshock ions that partially gyrate against the induced electric field in the solar wind rest frame.

Based on this model, a critical point of foreshock transient formation is the initiation of the "instability". For example, if the field variation by the Hall current cannot trap more foreshock ions to enhance the Hall current, a stable solution can be reached (see derivation in Appendix A). In this case, there is only static modification of the magnetic field profile around the discontinuity and no foreshock transient will form and grow. Here we consider two examples (Figure 10) with a B-field discontinuity that varies the IMF direction by positive and negative 90°, respectively. For simplicity, we ignore the thermal speed of foreshock ions so that the Hall current is only driven by the bulk velocity. Based on our model, when foreshock ions encounter the discontinuity, the Hall current can form in the direction mainly along the purple arrow. Such Hall current causes $+\Delta B_z$ near the discontinuity and $-\Delta B_z$ away from the discontinuity. In example 1 with $-B_z$ upstream of the discontinuity (Figure 10a), the Hall current decreases the field strength at the discontinuity meaning that foreshock ions can cross the discontinuity more easily resulting in stronger Hall current (instability). In example 2 with $+B_z$ upstream of the discontinuity (Figure 10b), the Hall current increases the field strength at the discontinuity meaning that less foreshock ions can cross the discontinuity resulting in weaker Hall current (stable solution). Interestingly, if the B-field discontinuity is a solar wind TD, the convection electric field points towards the TD in example 1 and points away in example 2 (green arrow). The convection electric field pointing towards the TD is one important characteristic of HFAs (e.g., Thomsen et al., 1993; Schwartz et al., 2000). Our model could at least partially explain this characteristic. As solar wind velocity is always earthward, the convection electric field direction indicates the magnetic field configuration relative to the TD. Such magnetic field configuration could determine whether the Hall current from foreshock ions can trigger the "instability".

Besides HFAs, SHFAs, and FBs, our model may also help explain the formation of other types of foreshock transients. For example, when a solar wind discontinuity separates the foreshock and the pristine solar wind and the discontinuity cannot trigger the "instability", a stable solution could be a local field strength enhancement at the discontinuity due to the foreshock ion Hall current (e.g., Figure 10b). Such kind of structure could be identified as a foreshock compressional boundary. Additionally, in event 2, the steepening process of the compressional boundary may also contribute to the steepening process of SLAMS. In the future, more case studies in comparison with simulations are needed to further investigate the formation and development process of various types of foreshock transients.

Our model only considers background foreshock ions as a beam based on observations (Figures 4 and 8). Due to the specular reflection at the bow shock, however, foreshock ions can also be gyrating ions with a certain gyrophase (e.g., Fuselier, 1995). Previous models (e.g., Burgess and Schwartz, 1988; Burgess 1989) show that specularly reflected ions can more easily channel along a TD when the convection electric field points towards it, which favors the formation of HFAs. In our model, what we concern is how specularly reflected ions contribute to the Hall current when they gyrate across a TD. Compared to a field-aligned beam, specularly reflected ions also have field-aligned motion and the difference is the large gyration with a certain gyrophase. Based on the model by Liu et al. (2015), when the convection electric field points towards the TD, the corresponding IMF configuration makes a single ion (similar to gyrophase bunched ions) prefer to project its initial velocity to the perpendicular direction, which favors strong Hall current. The Hall current direction, however, strongly depends on the initial gyrophase. This significantly complicates our model, which requires further study in the future.

Next, we discuss what may happen after the formation process. Based on our model, the energy source is foreshock ions. If foreshock ions cannot provide enough energy, the foreshock transient structure cannot be maintained and will become dissipated. For example, in the observations of "mature HFAs" (e.g., Zhang et al., 2010), foreshock ions and solar wind ions merged into one diffuse ion population meaning that there was no free energy from foreshock ions anymore. As a result, the Hall current strength decreased, so the magnetic field structure should become more gradual or less steepened. The induced electric field during this process should drive plasma moving inward (back to the core). Indeed, as observed by THEMIS when in a string-of-pearls formation, the density in the core of an FB increased during its late stage (Liu et al., 2016b).

As the energy source is foreshock ions, higher background foreshock ion density and energy must favor the formation process. As shown in PIC simulations (An et al., 2020 ApJ accepted), the expansion speed of a foreshock transient is proportional to normalized foreshock ion energy and the density ratio of foreshock ions to the solar wind ions. This density ratio is proportional to the Alfvén Mach number (see review by Lembege et al. (2004)). The foreshock ion speed is proportional to the solar wind speed, and the foreshock ion speed normalized to the Alfvén speed is also proportional to the Alfvén Mach number. This is consistent with previous statistical study (e.g., Liu et al., 2017a; Chu et al., 2017) and multi-case study (Liu et al., 2016; Turner et al., 2020) that fast solar wind speed and small field strength favor the occurrence of SHFAs/HFAs/FBs, and the expansion speed of FBs is proportional to the solar wind speed and the Alfvén Mach number.

Here we estimate the energy and momentum transfer from foreshock ions to the magnetic field and cold plasma. In event 2, the foreshock ion density at MMS2 $n_f$ was ~0.5-2 cm$^{-3}$ (Figure 9.1h), and the foreshock ion velocity $\boldsymbol{V_f}$ was ~500 km/s (Figure 8). The electric field was ~1 mV/m (Figure 9c). The energy transfer rate was thus around 0.04-0.16 nW/m$^3$. The acceleration of cold plasma expansion speed $v_{exp}$ was from ~0 to 50 km/s within 0.4s (Figure 8 from MMS2 to MMS1). The solar wind density at MMS2 $n_{sw}$ was ~2-10 cm$^{-3}$ (Figure 9.1h). The energy increase of cold plasma per unit time was $m \cdot n_{sw} \cdot v_{exp} \cdot \Delta v_{exp}/\Delta t$ ~ 0.02-0.1 nW/m$^3$, comparable to the energy input from foreshock ions. The magnetic energy also redistributed with a rate varied from -0.01 to +0.1 nW/m$^3$. As for the momentum, the transfer from foreshock ions to cold plasma was not straightforward, because there was no collision. In event 1, for example, the momentum of foreshock ions was mainly in the negative GSE-Y direction, but the outward motion of cold plasma was dominant in the GSE-XZ plane (Figure 3h). The momentum transfer was through the electric field and magnetic field $qn_f(\boldsymbol{E} + \boldsymbol{V_f} \times \boldsymbol{B})$. In event 2 at MMS2, $E_x \sim 0$ to -1 mV/m (Figure 9.1c) and $(\boldsymbol{V_f} \times \boldsymbol{B})_x \sim$ -0.5 to -4 mV/m which gives momentum input around $4\times10^{-17}$ to $1.6\times10^{-15}$ N/m³. The momentum gain of cold plasma per unit time was $m \cdot n_{sw} \cdot \Delta v_{exp}/\Delta t \sim 4\times10^{-16}$ to $2\times10^{-15}$ N/m³ (likely overestimated as there was also earthward expansion at the leading edge seen in Figure 9f), which was comparable to the momentum input from foreshock ions to the field. The momentum of magnetic field also redistributed, which was too complicated to be estimated.

In both events, no clear downstream compressional boundary was observed. This could be due to the direction of the Hall current. In event 1, as the initial bulk velocity of downstream foreshock ions was the dominant contribution to the Hall current, the Hall current was roughly along the downstream IMF direction. The field variation by the Hall current was thus roughly perpendicular to the downstream IMF (Figure 2b), which can hardly increase the field strength but mainly rotate the field direction. Therefore, the direction and strength of the Hall current and IMF configuration determine whether a compressional boundary can appear. Additionally, as there was magnetic flux transporting downstream, although the flux direction was roughly perpendicular to the downstream IMF, betatron acceleration could still occur which may explain the observed electron temperature increase at the downstream edge of event 1 (especially the perpendicular temperature in Figure 1f). This temperature increase was even higher than that at the upstream compressional boundary. A possible reason could be that electrons accelerated at the upstream compressional boundary could move along the same field lines towards the downstream edge (as solar wind electrons had earthward preference) and experience the second instance of betatron acceleration consistent with Liu et al. (2019). (We do not explain event 2, as we do not have good data to determine the discontinuity orientation and IMF configuration.)

Based on our model, we discuss what might cause the differences between HFAs and FBs that FBs are typically larger than HFAs with a shock at the upstream side. If foreshock ions from the downstream side can cross the discontinuity to the upstream side (rotational discontinuity or if the TD thickness is small enough compared to foreshock ion gyroradius), and the upstream IMF configuration can trigger the "instability", a bump can form upstream of a tenuous core. If the energy source from foreshock ions is strong enough, the expansion can be supermagnetosonic and the upstream bump can thus steepen into a shock. Also because of the supermagnetosonic expansion, the size of the core will soon become very large. We will identify such a structure as an FB. In some other cases, for example, if the expansion is very slow, no shocks will form, and the structure size will be small. If both sides could trigger the "instability" or the downstream magnetic flux transport is strong enough to increase the field strength, there could be two compressional boundaries or shocks on two sides of the core. Or if the TD is too thick, a local structure will form. We may identify those structures as HFAs. Although HFAs and FBs share similarities in the formation process and observational characteristics, we still need to distinguish them, because the effects of FBs are more significant. For example, because of the large size, the perturbations on the bow shock and magnetosphere/ionosphere by an FB can be global compared to typical HFAs (e.g., Archer et al., 2015). Because of the upstream shock and no significant downstream compression region, FBs can accelerate particles more significantly than many HFAs that do not have shocks, e.g., through shock drift acceleration and Fermi acceleration (Liu et al., 2016a; Liu et al., 2017b; Liu et al., 2018).

Foreshock transients have been shown to contribute to particle acceleration at Earth's bow shock. Foreshock transients potentially contribute to the energy budget at other astrophysical shocks, like supernova-driven shocks, but direct observations are unavailable for those more exotic systems. Our model sheds light on the quantification of the formation of foreshock transients to infer whether they can form at other shock environments to include them in general shock models. For example, our results imply that at shocks with an Alfvén Mach number larger than Earth's bow shock, foreshock transients could be more significant and occur more frequently. In the future, theoretical work and simulations can be applied to improve and refine our model.

**Figures**

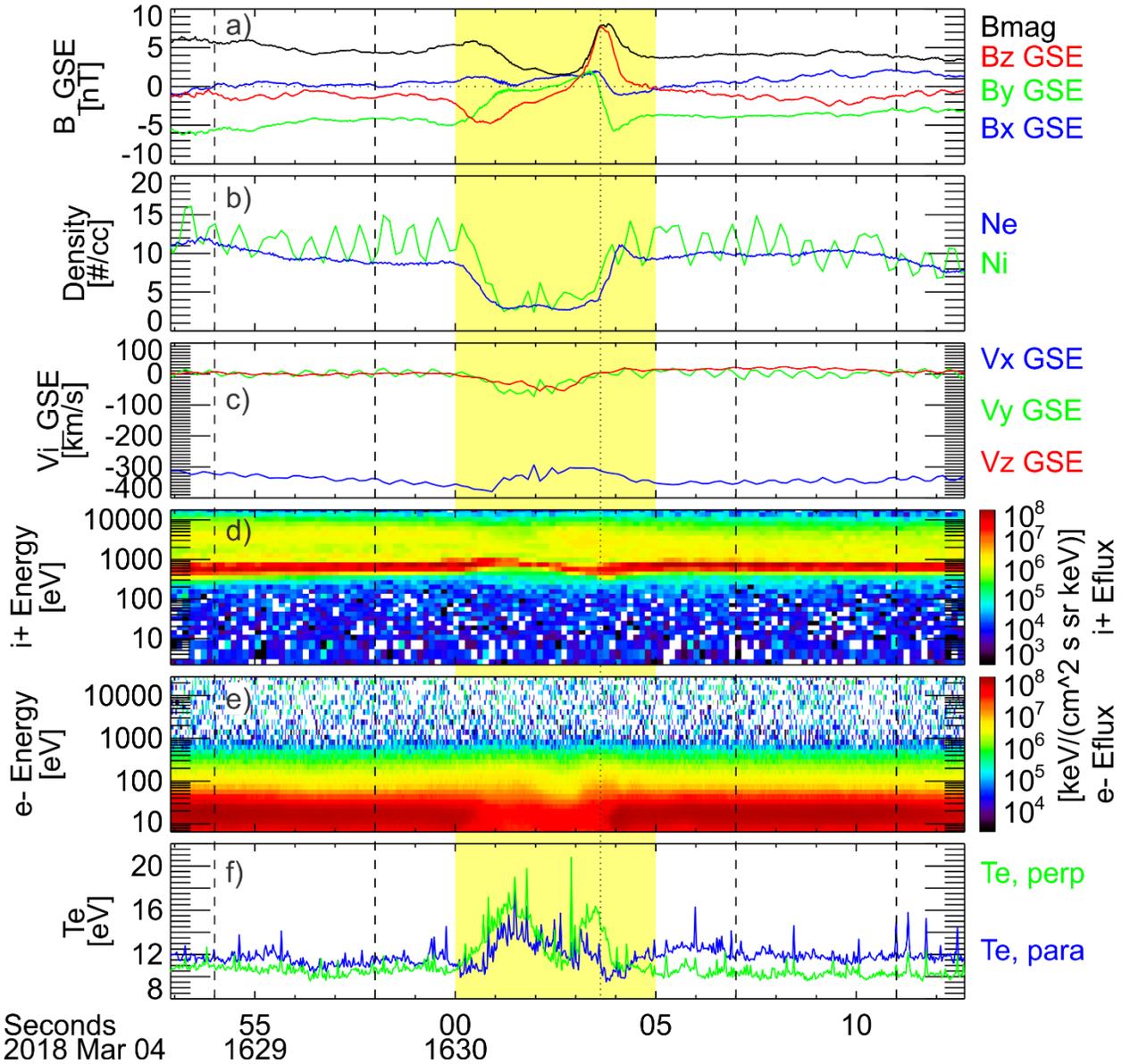

**Figure 1.** MMS1 observations of event 1. From top to bottom are: (a) magnetic field in GSE; (b) plasma density; (c) ion bulk velocity in GSE; (d) ion energy flux spectrum; (e) electron energy flux spectrum; (f) electron parallel (blue) and perpendicular (green) temperature. Vertical dashed lines indicate the time interval used to calculate the TD normal. Vertical dotted line indicates the compressional boundary. The yellow shaded region marks the HFA/FB.

a) overview

b) details in the solar wind rest frame

**Figure 2.** The sketch of event 1. (a) the overall geometry of the event. The event was observed at the flank of the bow shock with IMF (blue arrows) dominant in the negative GSE-Y direction. The convection electric field (green arrows) upstream of the possible TD pointed towards the TD. The TD normal was mainly in the GSE-Z direction and the upstream compressional boundary (gray box) was mainly sunward. (b) the zoom-in sketch in the XZ plane corresponding to the pink box in (a). In the solar wind rest frame, as MMS crossed the event (gray dashed arrow), it observed curved magnetic field lines (blue) caused by the Hall current (purple) and the corresponding outward plasma flow speed (red arrows) driven by the electric field (green). Foreshock ions were mainly moving in the negative GSE-Y direction (orange out-of-plane symbol) with gyration in the GSE-XZ plane (orange arrow).

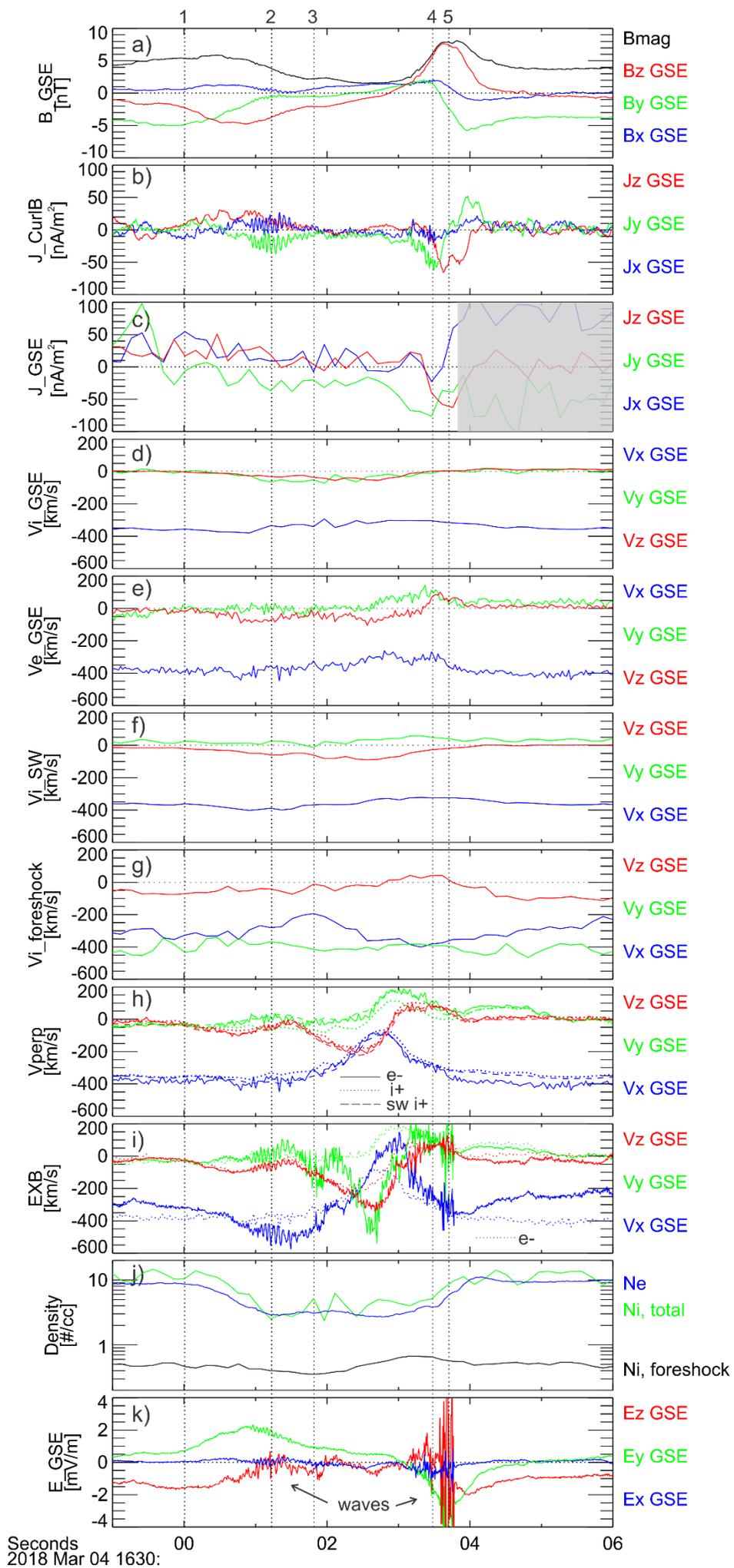

**Figure 3.** Current density configuration of event 1. From top to bottom are: (a) magnetic field; (b) current density calculated from the curlometer method; (c) current density calculated from plasma data (inaccurate in the gray shaded region likely due to the large measurement uncertainty of narrow solar wind ion beam); (d) total ion bulk velocity; (e) electron bulk velocity; (f) solar wind ion bulk velocity; (g) foreshock ion bulk velocity; (h) perpendicular velocity of electrons (solid), solar wind ions (dashed), and total ions (dotted); (i) E×B drift velocity in comparison with electron perpendicular velocity (dotted), and their differences may be due to measurement uncertainty of electric field in the spin axis direction and other drifts, such as diamagnetic drift; (j) total plasma density and foreshock ion density; (k) electric field interpolated to the magnetic field resolution (to better examine the DC electric field). The electric field large amplitude fluctuations in the compressional boundary were very likely whistlers triggered by electron perpendicular temperature anisotropy (Shi et al., 2020). Vertical dotted lines indicate the time of ion distributions in Figure 4.

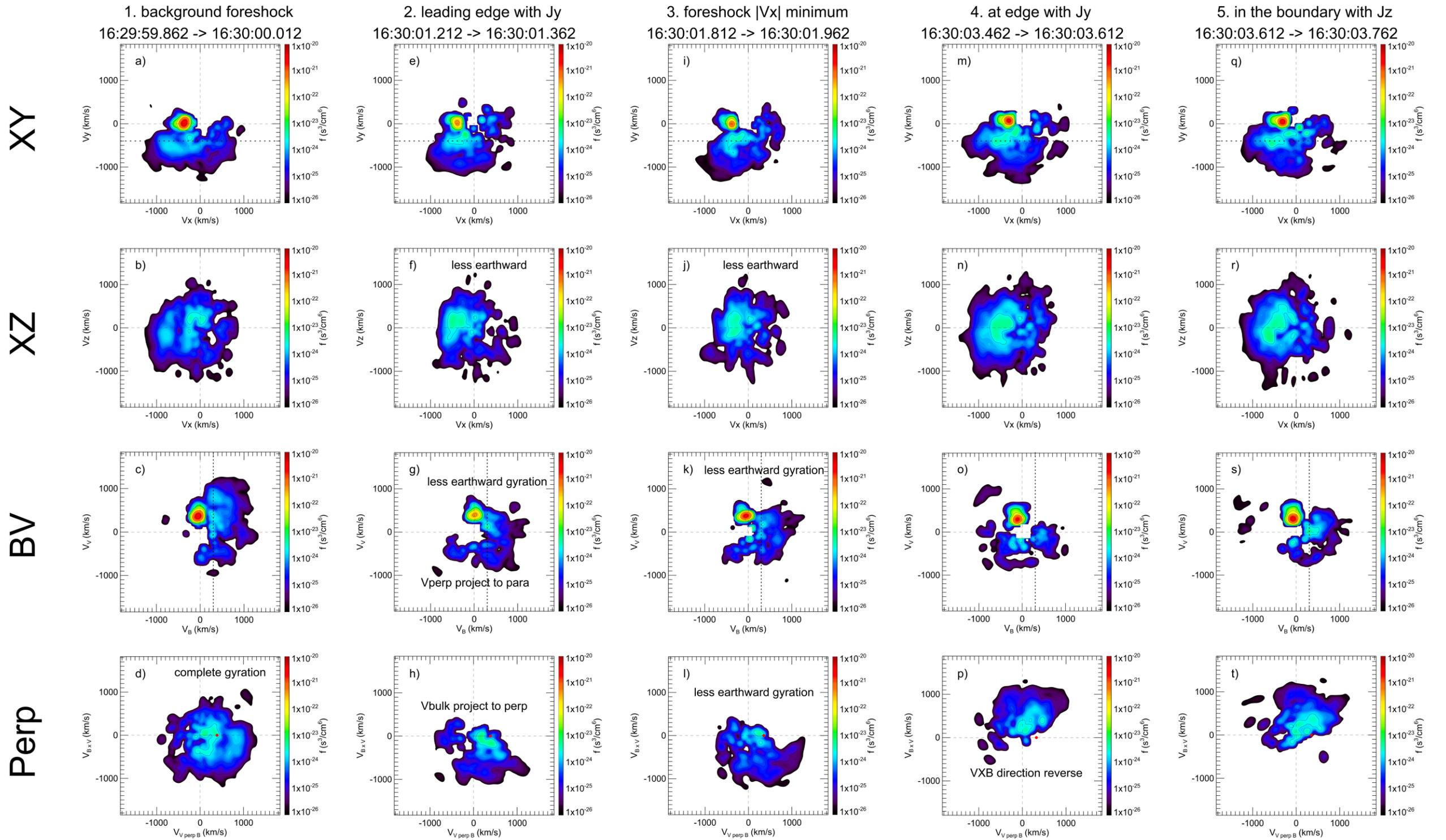

**Figure 4.** Ion distributions around the event. Each column corresponds to a vertical dotted line in Figure 3. From top to bottom are slices in GSE-XY plane at $V_z=0$ km/s, GSE-XZ plane at $V_y=-400$ km/s (cut through the horizontal dotted line in the XY plane), BV plane (horizontal axis is along the magnetic field direction and the plane contains the electron bulk velocity), perpendicular or gyrophase plane at $V_{para}=300$ km/s (cut through the vertical dotted line in the BV plane; horizontal axis is roughly along **E**×**B** and vertical axis is along **B**×**V**). The red dots in the perpendicular plane indicate the gyrocenter.

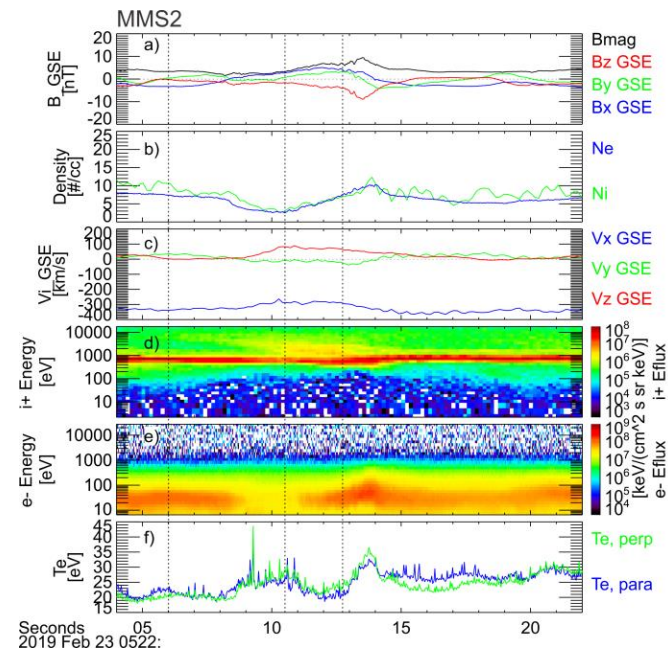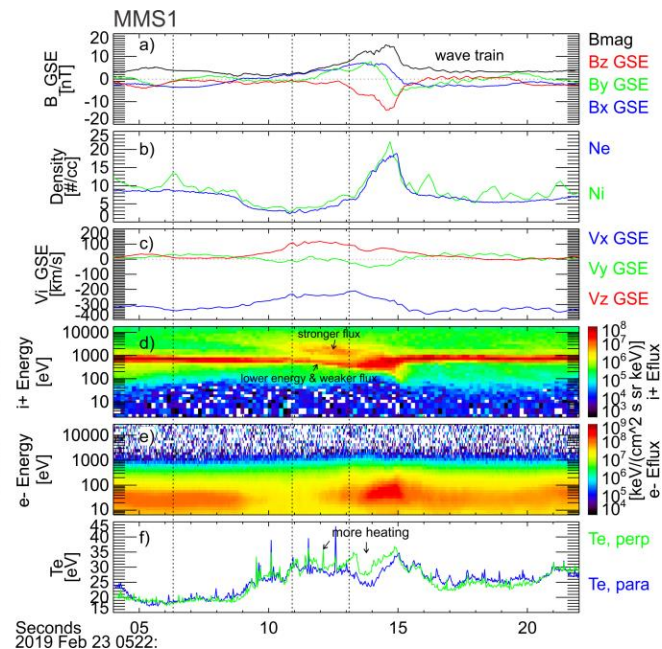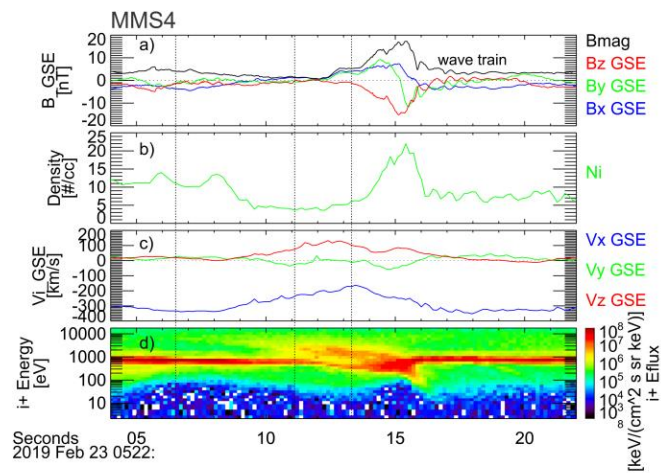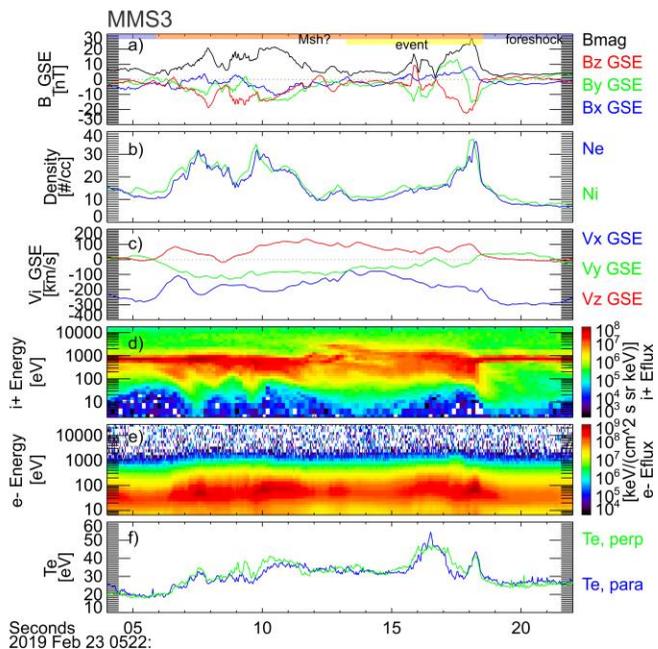

**Figure 5.** Observations of event 2 in the sequence of MMS2 (Figure 5.1), MMS1 (Figure 5.2), MMS4 (Figure 5.3), and MMS3 (Figure 5.4). From top to bottom are: (a) magnetic field; (b) density; (c) ion bulk velocity; (d) ion energy flux spectrum; (e) electron energy flux spectrum; (f) electron parallel (blue) and perpendicular (green) temperature. MMS4 FPI electron data was not available here. Vertical dotted lines in Figures 5.1-5.3 indicate the moments of ion distributions in Figure 8.

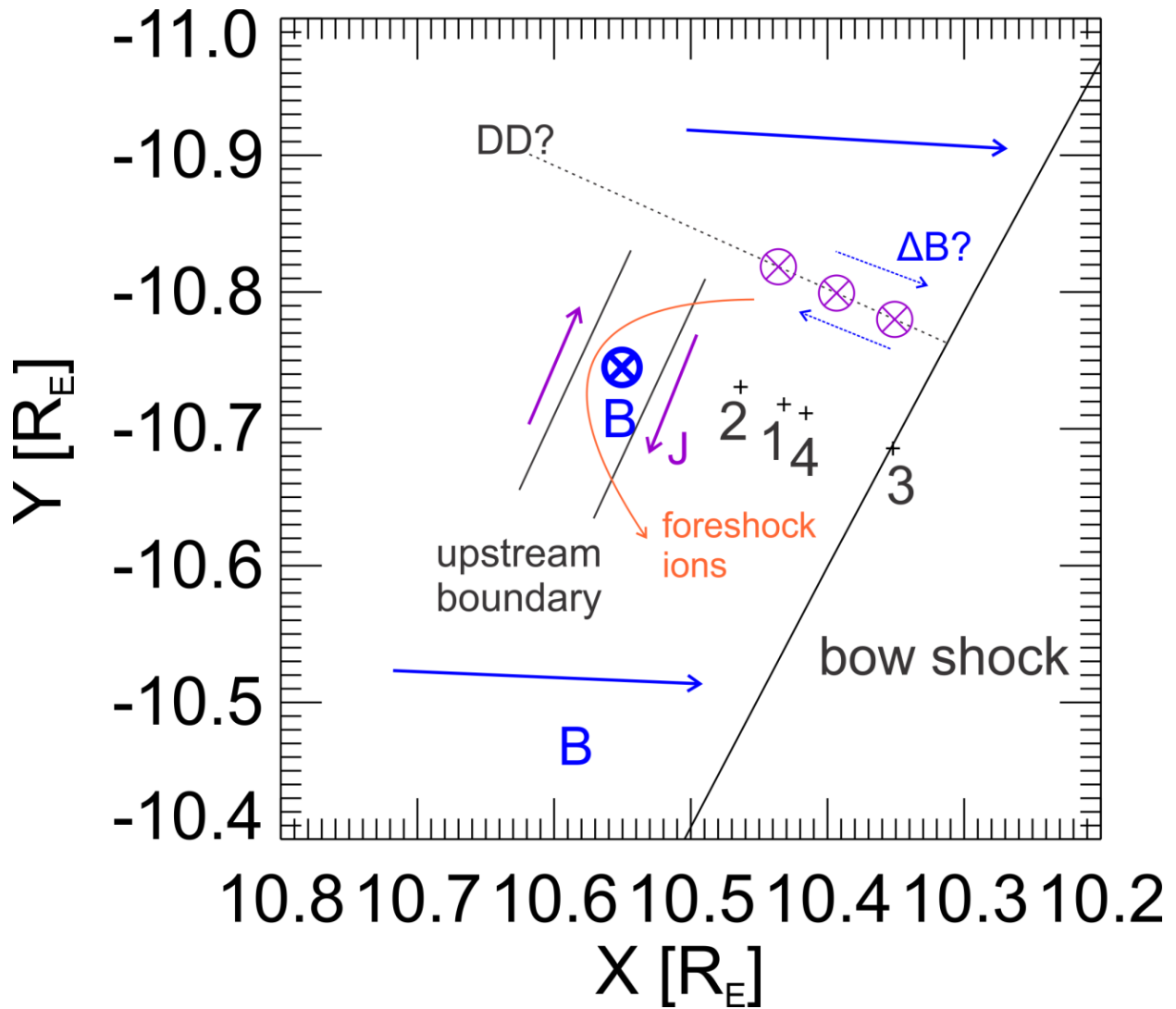

**Figure 6.** The geometry and spacecraft position in event 2. The IMF (blue) had strong radial component. The compressional boundary was mainly sunward (based on MVA results) with strong $B_z$ inside it (blue) and a bipolar current configuration (purple) on two sides. This compressional boundary reflected some sunward foreshock ions to be earthward (orange arrow). The position of B-field directional discontinuity (DD) was inferred from $B_x$ reversal and the corresponding $J_z$ observed at ~05:22:09 UT (Figure 9b).

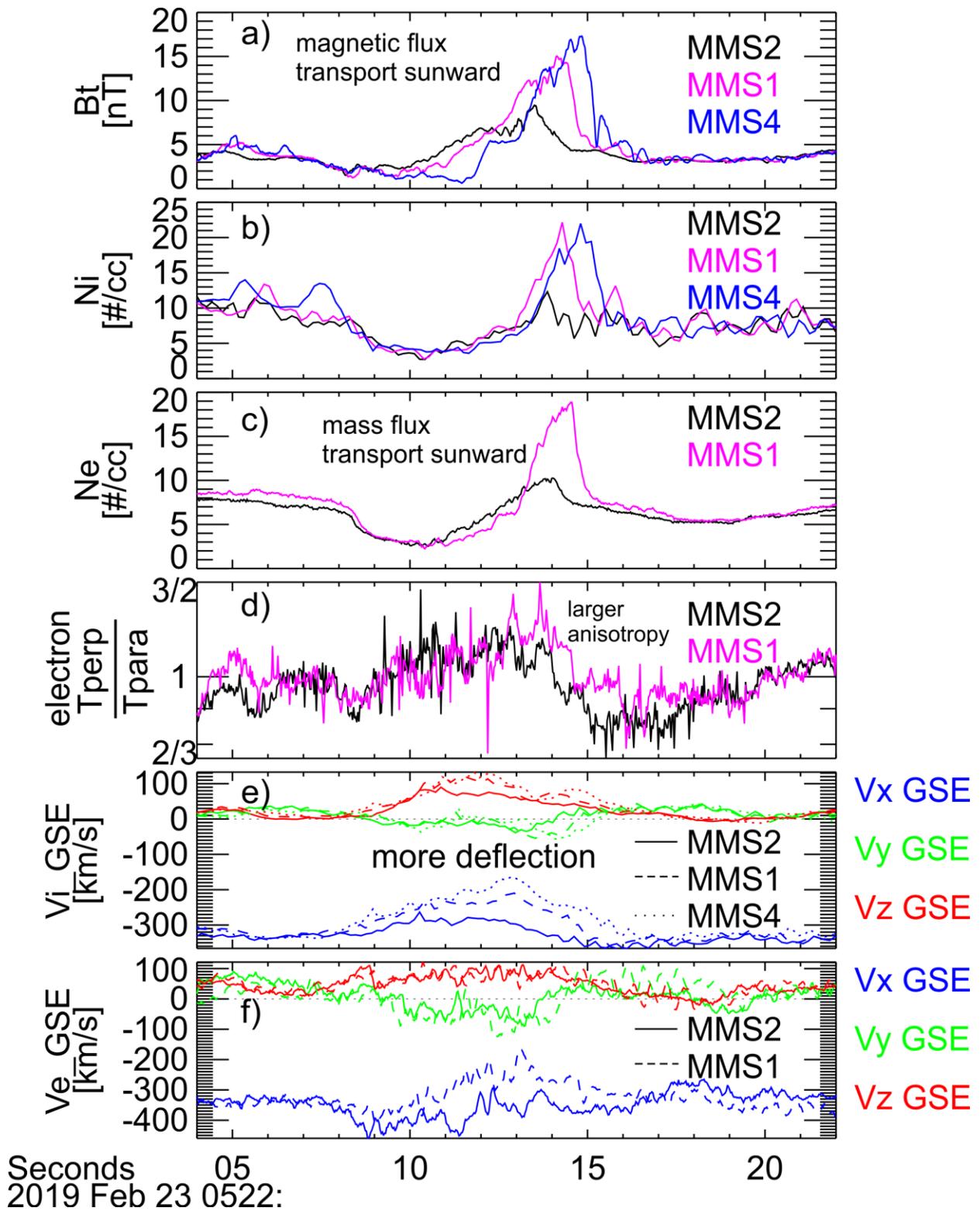

**Figure 7.** Superposed plots of MMS2, MMS1, and MMS4. MMS1 and MMS4 were time shifted by -0.4s and -0.6s, respectively, so that three spacecraft observations start roughly at the same

time. From top to bottom are their comparison of: (a) magnetic field strength; (b) ion density; (c) electron density; (d) electron temperature ratio of perpendicular to parallel; (e) total ion bulk velocity; (f) electron bulk velocity.

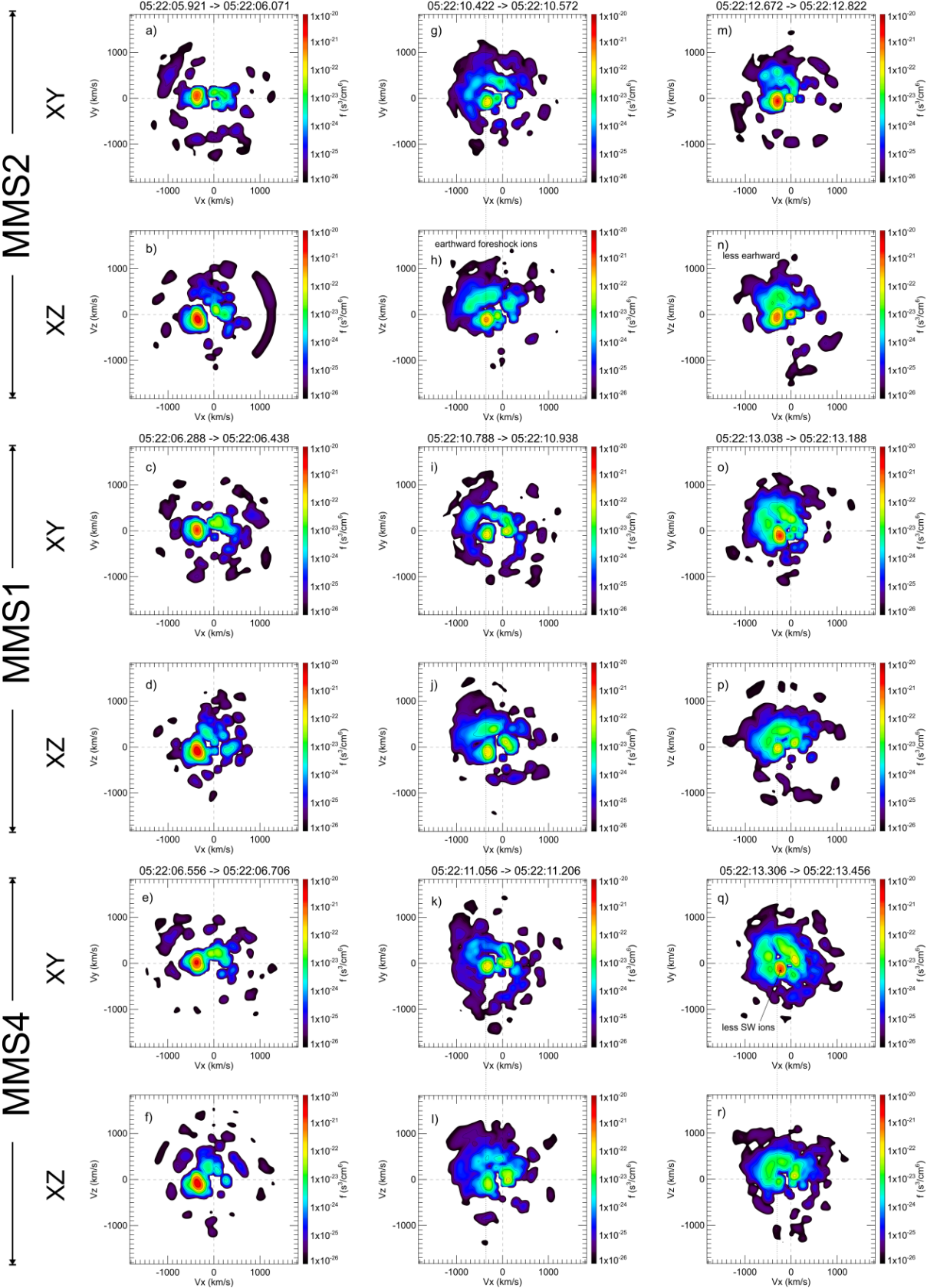

**Figure 8.** Ion distributions of MMS2, MMS1, and MMS4 in the GSE-XY and XZ slices at the moments corresponding to vertical dotted lines in Figures 5.1-5.3. There is ~0.4s and 0.6s time differences between distributions by MMS2 and MMS1 and by MMS2 and MMS4 in each column, respectively. The vertical dotted lines indicate the solar wind ion velocity measured by MMS2 to compare with MMS1 and MMS4 measurements.

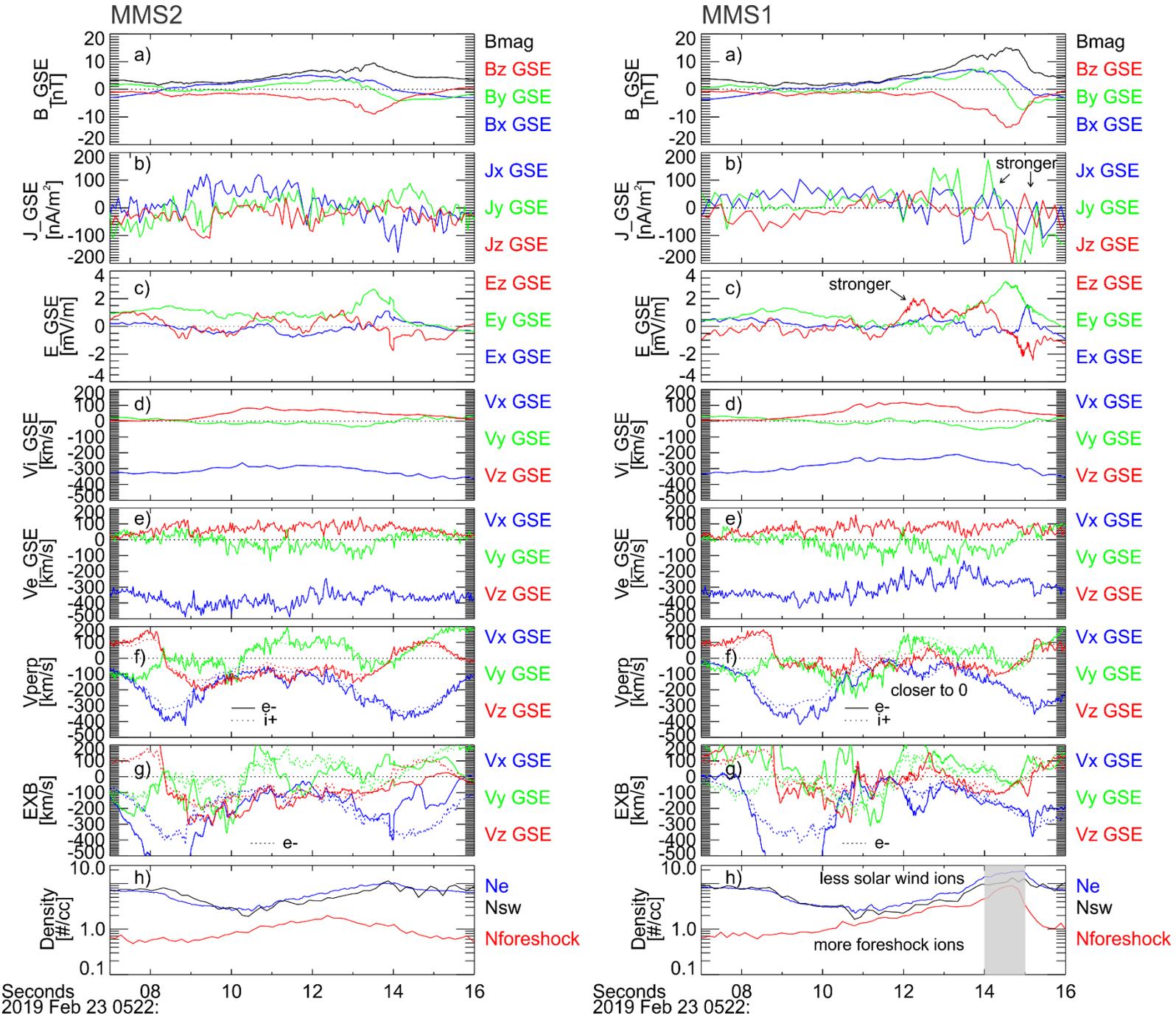

**Figure 9.** The evolution of the current density from MMS2 (Figure 9.1) to MMS1 (Figure 9.2). From top to bottom are: (a) magnetic field; (b) current density calculated from plasma data; (c) electric field interpolated to the magnetic field resolution; (d) total ion bulk velocity; (e) electron bulk velocity; (f) perpendicular velocity of electrons (solid) and total ions (dotted); (g) E×B

velocity in comparison with electron perpendicular velocity; (h) density of electrons (blue), solar wind ions (black), and foreshock ions (red). At MMS1, because the ion distribution was rather diffuse in the compressional boundary, the density of foreshock ions may be overestimated, so this part is gray shaded.

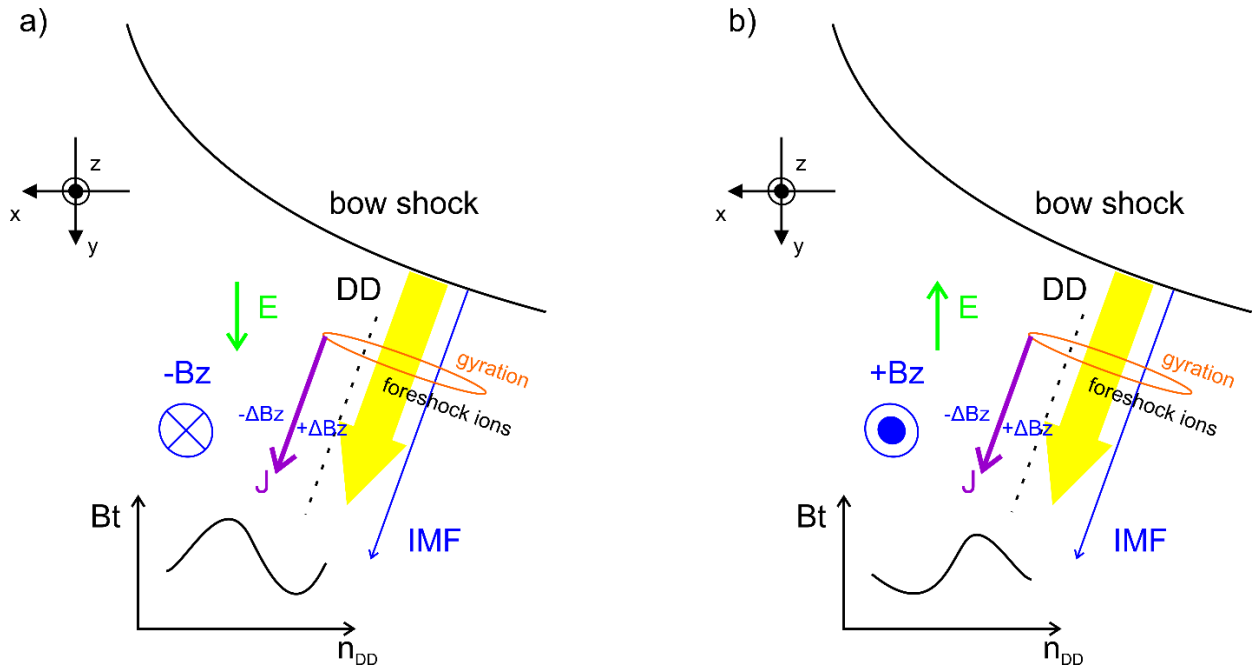

**Figure 10.** Two examples of IMF configuration. In both examples, the downstream IMF is along the positive GSE-X and Y direction. In example 1 (a), the upstream IMF is along the negative GSE-Z direction and the convection electric field points towards the B-field directional discontinuity (DD). Due to the Hall current, the field strength decreases at the DD (the small panel shows Bt variation along the DD normal). In example 2 (b), the upstream IMF and the convection electric field direction reverses. The Hall current increases the field strength at the DD.

**Acknowledgements**

We acknowledge the International Space Science Institute (ISSI) team "Resolving the Microphysics of Collisionless Shock Waves" led by L. Wilson for providing a collaborative opportunity for this work and the ISSI team "Foreshocks Across The Heliosphere: System Specific Or Universal Physical Processes?" led by H. Hietala & F. Plaschke for the meaningful discussion. T. Z. L. is supported by the NASA Living With a Star Jack Eddy Postdoctoral Fellowship Program, administered by the Cooperative Programs for the Advancement of Earth System Science (CPAESS). T. Z. L. and X. A. are partially supported by NSF award AGS-1941012. H. Z. acknowledges NSF AGS-1352669 and NASA contract 80NSSC18K1376. MMS data are available at MMS Science Data Center (https://lasp.colorado.edu/mms/sdc/). We thank the SPEDAS software team and NASA's Coordinated Data Analysis Web (CDAWeb, http://cdaweb.gsfc.nasa.gov/) for their analysis tools and data access. The SPEDAS software (see Angelopoulos et al. (2019)) is available at http://themis.ssl.berkeley.edu.

**Appendix A. A Simple Derivation**

Here we derive a growth rate of the "instability" based on our model. Considering a 1-D magnetic profile based on event 1, magnetic field is in Z direction with a profile varying in X direction. The Hall current from the projected perpendicular velocity of demagnetized foreshock ions is in Y direction. We ignore the motion of cold plasma in the solar wind rest frame (and thus the background convection electric field). For linearization, we have foreshock ion density $n_f = n_{f0} + n_{f1}$, foreshock ion velocity $\boldsymbol{V_f} = V_{f0}\hat{\boldsymbol{y}} + \boldsymbol{V_{f1}}$, magnetic field $\boldsymbol{B} = B_0\hat{\boldsymbol{z}} + \boldsymbol{B_1}$, electric field $\boldsymbol{E} = \boldsymbol{E_1}$, $\frac{\partial}{\partial t} \to -i\omega$, and $\nabla \to ik\hat{\boldsymbol{x}}$.

Foreshock ion density continuity equation: $\frac{\partial n_f}{\partial t} + \nabla \cdot (n_f \mathbf{V}_f) = \frac{\delta n_f}{\delta t}$, where $\frac{\delta n_f}{\delta t}$ indicates the particle source from the background foreshock. In events 1 and 2, we show that stronger compressional boundary can trap more foreshock ions at the magnetic gradient by limiting their gyration (Figures 3j and 9h). We thus assume that $\frac{\delta n_f}{n_f} = i\alpha \frac{B_1}{B_0}$, where $\alpha$ is a parameter in a scale of 1 simplifying how the magnetic profile variation traps foreshock ions. The sign of $\alpha$ indicates whether the magnetic profile variation traps more foreshock ions. If the magnetic profile variation increases the field strength at the compressional boundary (Figure 10a) to trap more foreshock ions, $\alpha$ will be positive. If the magnetic profile variation increases the field strength at the discontinuity (Figure 10b), which causes less foreshock ions to pass through the discontinuity and contribute to the Hall current, $\alpha$ will be negative. At the end of Appendix, we show that the sign of $\alpha$ is critical to determine whether the "instability" can occur. From Figures 3j and 9h, we see that the foreshock ions were accumulated at the magnetic gradient, we thus include $i$ to indicate 90° phase difference in $k$ between the peak of foreshock ion density and magnetic field strength. To linearize the continuity equation, we have $-i\omega n_{f1} + ikn_{f0}V_{f1x} = -i\omega(i\alpha \frac{B_1}{B_0} n_{f0})$, and thus $n_{f1} = \frac{k}{\omega} n_{f0} V_{f1x} + i\alpha \frac{B_1}{B_0} n_{f0}$ (A1).

Next, foreshock ion momentum equation: $m \frac{d\mathbf{V}_f}{dt} = e\mathbf{E} + e\mathbf{V}_f \times \mathbf{B}$. Ignoring the variation in $V_{f0}$, we have $-i\omega m \mathbf{V}_{f1} = e\mathbf{E}_1 + e\mathbf{V}_{f1} \times \mathbf{B}_0 + e\mathbf{V}_{f0} \times \mathbf{B}_1$ (A2). From Faraday's Law: $i\mathbf{k} \times \mathbf{E}_1 = i\omega \mathbf{B}_1$. If $\mathbf{B}_1$ is only in Z direction, i.e., we only consider field strength variation, we simplify it as $ikE_1 = i\omega B_1$ and $\mathbf{E}_1$ is only in Y direction. Eq. (A2) can thus be simplified as: $-i\omega m V_{f1x} = eV_{f1y}B_0 + eV_{f0}B_1$ and $-i\omega m V_{f1y} = eE_1 - eV_{f1x}B_0$. Solving for $V_{f1x}$ and $V_{f1y}$, we have $V_{f1x} = \frac{\Omega}{\omega} eE_1/\omega m \left(1 - \frac{\Omega^2}{\omega^2}\right) - eV_{f0}B_1/i\omega m \left(1 - \frac{\Omega^2}{\omega^2}\right)$ and $V_{f1y} = $

$-eE_1/i\omega m\left(1-\frac{\Omega^2}{\omega^2}\right)+\frac{\Omega}{\omega}eV_{f0}B_1/\omega m\left(1-\frac{\Omega^2}{\omega^2}\right)$, where $\Omega$ is the ion gyrofrequency. Because as seen from event 2 the field variation is much faster than the ion gyrofrequency ($\frac{\Omega}{\omega}\ll 1$), we simplify them as $V_{f1x}=-eV_{f0}B_1/i\omega m$ and $V_{f1y}=-eE_1/i\omega m$ (A3).

From Ampere's Law: $i\mathbf{k}\times\mathbf{B_1}=\mu_0\mathbf{J_1}$. As we only consider field strength variation, we have $-ikB_1=\mu_0 J_1$. The current density variation $J_1=en_{f1}V_{f0}+en_{f0}V_{f1y}$. Using Eq. (A1) and (A3), we have $J_1=e\left(\frac{k}{\omega}n_{f0}V_{f1x}+i\alpha\frac{B_1}{B_0}n_{f0}\right)V_{f0}-e^2n_{f0}E_1/i\omega m=e\left(\frac{k}{\omega}n_{f0}(-eV_{f0}B_1/i\omega m)+i\alpha\frac{B_1}{B_0}n_{f0}\right)V_{f0}-e^2n_{f0}E_1/i\omega m=-e^2\left(\frac{k}{\omega}\right)^2n_{f0}V_{f0}^2E_1/i\omega m+i\alpha e\frac{k}{\omega}\frac{E_1}{B_0}n_{f0}V_{f0}-e^2n_{f0}E_1/i\omega m$. We thus have $-ikB_1=-i\frac{k^2}{\omega}E_1=\mu_0 J_1=-\frac{\mu_0 e^2 n_{f0}}{i\omega m}\left(1+\left(\frac{k}{\omega}V_{f0}\right)^2\right)E_1+i\mu_0\alpha e\frac{k}{\omega}\frac{E_1}{B_0}n_{f0}V_{f0}$ and get $k^2=-\frac{\omega_{pf}^2}{c^2}\left(1+\left(\frac{k}{\omega}V_{f0}\right)^2\right)-\alpha\frac{\omega_{pf}^2}{c^2}\frac{kV_{f0}}{\Omega}$. Finally, we have

$$\omega^2=-(kV_{f0})^2/(1+k^2c^2/\omega_{pf}^2+\alpha\,kV_{f0}/\Omega) \qquad (A4)$$

It is a modification of foreshock ion beam instability by involving a particle source that is modulated by the magnetic variation. We see that if $\alpha>0$, $\omega^2$ is always negative meaning instability. Based on observations, we estimate $k\sim 2\pi/2000\,km^{-1}$, $\frac{c}{\omega_{pf}}\sim 300\,km$, $V_{f0}\sim 500\,km/s$, and $\Omega\sim 2\pi/15\,s^{-1}$. Therefore, $\omega^2\sim -14/(1.9+3.8\alpha)\,s^{-2}$. If $\alpha\sim 1$, we obtain a growth rate of $\sim 2.4\,s^{-1}$. Such a growth rate is consistent with the time scale of field variation in event 2. If the magnetic field variation traps less foreshock ions (Figure 10b) and $\alpha\sim -1$, $\omega^2>0$ meaning no instability. In this simplified derivation, many processes are ignored, such as static electric field, electron Hall current, and diamagnetic current. In the future, more comprehensive theoretical work is needed.

**References**

An, X., T. Z. Liu, J. Bortnik, A. Osmane, V. Angelopoulos (2020). Formation of foreshock transients and associated secondary shocks. Accepted by ApJ. https://arxiv.org/abs/2008.03245

Angelopoulos, V., Cruce, P., Drozdov, A. et al. Space Sci Rev (2019) 215: 9. https://doi.org/10.1007/s11214-018-0576-4

Archer, M. O., D. L. Turner, J. P. Eastwood, T. S. Horbury, and S. J. Schwartz (2014), The role of pressure gradients in driving sunward magnetosheath flows and magnetopause motion, J. Geophys. Res. Space Physics, 119, 8117–8125, doi:10.1002/2014JA020342.

Archer, M. O., D. L. Turner, J. P. Eastwood, S. J. Schwartz, and T. S. Horbury (2015), Global impacts of a Foreshock Bubble: Magnetosheath, magnetopause and ground-based observations, Planet. Space Sci., 106, 56-65, doi:10.1016/j.pss.2014.11.026

Bai, S.-C., Shi, Q., Liu, T. Z., Zhang, H., Yue, C., Sun, W.-J., et al. (2020). Ion-scale flux rope observed inside a hot flow anomaly. Geophysical Research Letters, 47, e2019GL085933. https://doi.org/10.1029/2019GL085933

Blanco-Cano, X., P. Kajdič, N. Omidi, and C. T. Russell (2011), Foreshock cavitons for different interplanetary magnetic field geometries: Simulations and observations, J. Geophys. Res., 116, A09101, doi:10.1029/2010JA016413.

Burch, J. L., Moore, T. E., Torbert, R. B., & Giles, B. L. (2016). Magnetospheric multiscale overview and science objectives. Space Science Reviews, 199(1 - 4), 5 – 21. https://doi.org/10.1007/s11214-015-0164-9

Burgess, D., & Schwartz, S. J. (1988). Colliding plasma structures: Current sheet and perpendicular shock. Journal of Geophysical Research, 93, 11,327–11,340. https://doi.org/10.1029/JA093iA10p11327

Burgess, D. (1989). On the effect of a tangential discontinuity on ions specularly reflected at an oblique shock. Journal of Geophysical Research, 94, 472–478. https://doi.org/10.1029/JA094iA01p00472

Chu, C., H. Zhang, D. Sibeck, A. Otto, Q. Zong, N. Omidi, J. P. McFadden, D. Fruehauff, and V. Angelopoulos (2017), THEMIS satellite observations of hot flow anomalies at Earth's bow shock, Ann. Geophys., 35, 3, 443–451, doi:10.5194/angeo-35-443-2017.

Eastwood, J. P., E. A. Lucek, C. Mazelle, K. Meziane, Y. Narita, J. Pickett, and R. A. Treumann (2005), The Foreshock, Space. Sci. Rev., 118, 41–94, https://doi.org/10.1007/s11214-005-3824-3.

Ergun, R. E., Tucker, S., Westfall, J., Goodrich, K. A., Malaspina, D. M., Summers, D.,…Cully, C. M. (2016). The axial double probe and fields signal processing for the MMS mission. Space Science Reviews, 199, 167–188.

Fuselier, S. A. (1995). Ion distributions in the Earth's foreshock upstream from the bow shock. Adv. Space Res. 15, 43–52. doi:10.1016/0273-1177(94)00083-D

Hartinger, M. D., D. L. Turner, F. Plaschke, V. Angelopoulos, and H. Singer (2013), The role of transient ion foreshock phenomena in driving Pc5 ULF wave activity, J. Geophys. Res. Space Physics, 118, doi:10.1029/2012JA018349.

Kis, A., O. Agapitov, V. Krasnoselskikh, Y. V. Khotyaintsev, I. Dandouras, I. Lemperger, and V. Wesztergom (2013), Gyrosurfing acceleration of ions in front of Earth's quasi-parallel bow shock, Astrophys. J., 771, 4, doi:10.1088/0004-637X/771/1/4.

Le Contel, O., Leroy, P., Roux, A., Coillot, C., Alison, D., Bouabdellah, A., et al. (2016). The search‐coil magnetometer for MMS. Space Science Reviews, 199(1‐4), 257–282. https://doi.org/10.1007/s11214-014-0096-9

Lembege, B., J. Giacalone, M. Scholer, T. Hada, M. Hoshino, V. Krasnosel'skikh, H. Kucharek, P. Savoini, and T. Terasawa (2004), Selected problems in collisionless‐shock physics, Space Sci. Rev., 110(3/4), 161–226, doi:10.1023/B:SPAC.0000023372.12232.b7.

Lin, Y. (1997), Generation of anomalous flows near the bow shock by its interaction with interplanetary discontinuities, J. Geophys. Res., 102 (A11), 24265–24281, doi:10.1029/97JA01989.

Lin, Y. (2002). Global hybrid simulation of hot flow anomalies near the bow shock and in the magnetosheath. Planet. Space Sci. 50, 577–591, doi:10.1016/S0032-0633(02)00037-5

Lin, Y. (2003), Global-scale simulation of foreshock structures at the quasi-parallel bow shock, J. Geophys. Res., 108, 1390, doi:10.1029/2003JA009991, A11.

Lindqvist, P.-A., Olsson, G., Torbert, R. B., King, B., Granoff, M., Rau, D., et al. (2016). The spin-plane double probe electric field instrument for MMS. Space Science Reviews, 199(1-4), 137–165. https://doi.org/10.1007/s11214-014-0116-9

Liu, Z., Turner, D. L., Angelopoulos, V., & Omidi, N. (2015). THEMIS observations of tangential discontinuity-driven foreshock bubbles. Geophysical Research Letters, 42, 7860–7866. https://doi.org/10.1002/2015GL065842

Liu, T. Z., Hietala, H., Angelopoulos, V., & Turner, D. L. (2016a). Observations of a new foreshock region upstream of a foreshock bubble's shock. Geophysical Research Letters, 43, 4708–4715. https://doi.org/10.1002/2016GL068984

Liu, T. Z., D. L. Turner, V. Angelopoulos, and N. Omidi (2016b), Multipoint observations of the structure and evolution of foreshock bubbles and their relation to hot flow anomalies, J. Geophys. Res. Space Physics, 121, doi:10.1002/2016JA022461.

Liu, T. Z., Angelopoulos, V., Hietala, H., & Wilson, L. B. III (2017a). Statistical study of particle acceleration in the core of foreshock transients. Journal of Geophysical Research: Space Physics, 122, 7197–7208. https://doi.org/10.1002/2017JA024043

Liu, T. Z., Lu, S., Angelopoulos, V., Hietala, H., & Wilson, L. B. III (2017b). Fermi acceleration of electrons inside foreshock transient cores. Journal of Geophysical Research: Space Physics, 122, 9248–9263. https://doi.org/10.1002/2017JA024480

Liu, T. Z., Lu, S., Angelopoulos, V., Lin, Y., & Wang, X. Y. (2018). Ion acceleration inside foreshock transients. Journal of Geophysical Research: Space Physics, 123, 163–178. https://doi.org/10.1002/2017JA024838

Liu, T. Z., Angelopoulos, V., and Lu, S. (2019), Relativistic electrons generated at Earth's quasi-parallel bow shock, Science Advances, 5, 7, doi:10.1126/sciadv.aaw1368

Liu, T. Z., Lu, S., Turner, D. L., Gingell, I., Angelopoulos, V., Zhang, H., et al. (2020). Magnetospheric Multiscale (MMS) observations of magnetic reconnection in foreshock transients. Journal of Geophysical Research: Space Physics, 125, e2020JA027822. https://doi.org/10.1029/2020JA027822

Omidi, N. and D. G. Sibeck (2007), Formation of hot flow anomalies and solitary shocks, J. Geophys. Res., 112, A10203, doi:10.1029/2006JA011663.

Omidi, N., J. P. Eastwood, and D. G. Sibeck (2010), Foreshock bubbles and their global magnetospheric impacts, J. Geophys. Res., 115, A06204, doi:10.1029/2009JA014828.

Omidi, N., H. Zhang, D. Sibeck, and D. Turner (2013), Spontaneous hot flow anomalies at quasi-parallel shocks: 2. Hybrid simulations, J. Geophys. Res., VOL. 118, 173–180, doi:10.1029/2012JA018099.

Pollock, C., Moore, T., Jacques, A., Burch, J., Gliese, U., Saito, Y., et al. (2016). Fast plasma investigation for magnetospheric multiscale. Space Science Reviews, 199(1 - 4), 331 – 406. https://doi.org/10.1007/s11214-016-0245-4

Robert, P., Dunlop, M. W., Roux, A., & Chanteur, G. (1998). Accuracy of current density determination. ISSI Scientific Reports Series, 1, 395–418.

Schwartz, S. J., et al., An active current sheet in the solar wind, Nature, 318, 269-271, 1985.

Schwartz, S. J., D. Burgess, W. P. Wilkinson, R. L. Kessel, M. Dunlop, and H. Luehr (1992), Observations of short large-amplitude magnetic structures at a quasi-parallel shock, J. Geophys. Res., 97, 4209–4227, doi:10.1029/91JA02581.

Schwartz, S. J. (1998), Shock and discontinuity normal, Mach numbers, and related parameters, from Analysis Methods for Multi-Spacecraft Data, edited by G. Paschmann and P. W. Daly, pp. 249–270.

Schwartz, S. J., Paschmann, G., Sckopke, N., Bauer, T. M., Dunlop, M., Fazakerley, A. N., and Thomsen, M. F. (2000), Conditions for the formation of hot flow anomalies at Earth's bow shock, J. Geophys. Res., 105(A6), 12639– 12650, doi:10.1029/1999JA000320.

Schwartz S. J., Avanov, L., Turner, D., Zhang, H., Gingell, I., Eastwood, J. P., et al. (2018). Ion kinetics in a hot flow anomaly: MMS observations. Geophysical Research Letters. 45. https://doi.org/10.1029/2018GL080189

Shi, X., Liu, T. Z., Angelopoulos, V., & Zhang, X. (2020). Whistler mode waves in the compressional boundary of foreshock transients. Journal of Geophysical Research: Space Physics, 125, e2019JA027758. https://doi.org/10.1029/2019JA027758

Sibeck, D. G., et al. (1999), Comprehensive study of the magnetospheric response to a hot flow anomaly, J. Geophys. Res., 104(A3), 4577–4593, doi:10.1029/1998JA900021.

Sibeck, D. G., T. D. Phan, R. P. Lin, R. P. Lepping, and A. Szabo (2002), Wind 556 observations of foreshock cavities: A case study, J. Geophys. Res., 107, 1271, 557 doi:10.1029/2001JA007539.

Sonnerup, B. U. Ö., & Scheible, M. (1998). Minimum and maximum variance analysis. In G. Paschmann & P. W. Daly (Eds.), Analysis methods for multi spacecraft data (pp. 185–215). Bern, Switzerland: European Space Agency.

Thomas, V. A., Winske, D., Thomsen, M. F., and Onsager, T. G. (1991), Hybrid simulation of the formation of a hot flow anomaly, J. Geophys. Res., 96(A7), 11625–11632, doi:10.1029/91JA01092.

Thomsen, M. F., Gosling, J. T., Fuselier, S. A., Bame, S. J., & Russell, C. T. (1986). Hot, diamagnetic cavities upstream from the Earth's bow shock. Journal of Geophysical Research, 91, 2961–2973. https://doi.org/10.1029/JA091iA03p02961

Thomsen, M. F., Gosling, J. T., Bame, S. J., Quest, K. B., & Russell, C. T. (1988). On the origin of hot diamagnetic cavities near the Earth's bow shock. Journal of Geophysical Research, 93, 11,311–11,325. https://doi.org/10.1029/JA093iA10p11311

Thomsen, M. F., Thomas, V. A., Winske, D., Gosling, J. T., Farris, M. H., & Russell, C. T. (1993). Observational test of hot flow anomaly formation by the interaction of a magnetic discontinuity with the bow shock. Journal of Geophysical Research, 98, 15,319–15,330. https://doi.org/10.1029/93JA00792

Treumann, R. A. (2009), Fundamentals of collisionless shocks for astrophysical application, 1. Non-relativistic shocks. Astron. Astrophys. Rev. 17, 409.

Turner D. L., N. Omidi, D. G. Sibeck, and V. Angelopoulos (2013), First observations of foreshock bubbles upstream of Earth's bow shock: Characteristics and comparisons to HFAs, J. Geophys. Res., VOL. 118, 1552–1570, doi:10.1002/jgra.50198.

Turner D. L. et al. (2018), Autogenous and efficient acceleration of energetic ions upstream of Earth's bow shock. Nature, 561, 206, doi:10.1038/s41586-018-0472-9

Turner, D. L., Liu, T. Z., Wilson, L. B., Cohen, I. J., Gershman, D. G., Fennell, J. F., et al (2020). Microscopic, multipoint characterization of foreshock bubbles with Magnetospheric Multiscale (MMS). Journal of Geophysical Research: Space Physics, 125, e2019JA027707. https://doi.org/10.1029/2019JA027707

Wang, B., Nishimura, Y., Hietala, H., Shen, X.-C., Shi, Q., Zhang, H., et al. (2018). Dayside magnetospheric and ionospheric responses to a foreshock transient on 25 June 2008: 2. 2-D evolution based on dayside auroral imaging. Journal of Geophysical Research: Space Physics, 123, 6347–6359. https://doi.org/10.1029/2017JA024846

Wang, S., et al., "Ion-scale Current Structures in Short Large Amplitude Magnetic Structures," Astrophys. J. 898(121), pp. 13, doi:10.3847/1538-4357/ab9b8b, 2020.

Wilson, L. B. III, A. Koval, D. G. Sibeck, A. Szabo, C. A. Cattell, J. C. Kasper, B. A. Maruca, M. Pulupa, C. S. Salem, and M. Wilber (2013), Shocklets, SLAMS, and field-aligned ion beams in the terrestrial foreshock, J. Geophys. Res. Space Physics, 118, 957–966, doi:10.1029/2012JA018186.

Wilson, L. B. I. I. I., Sibeck, D. G., Turner, D. L., Osmane, A., Caprioli, D., & Angelopoulos, V. (2016). Relativistic electrons produced by foreshock disturbances observed upstream of Earth's bow shock. Physical Review Letters, 117(21), 215101. https://doi.org/10.1103/PhysRevLett.117.215101

Wilson, L.B., III (2016). Low Frequency Waves at and Upstream of Collisionless Shocks. In Low‐Frequency Waves in Space Plasmas (eds A. Keiling, D.‐H. Lee and V. Nakariakov). doi:10.1002/9781119055006.ch16

Zhang, H., et al. (2010), Time history of events and macroscale interactions during substorms observations of a series of hot flow anomaly events, J. Geophys. Res., 115, A12235, doi:10.1029/2009JA015180.

Zhang H., D. G. Sibeck, Q.-G. Zong, N. Omidi, D. Turner, and L. B. N. Clausen (2013), Spontaneous hot flow anomalies at quasi-parallel shocks: 1. Observations, J. Geophys. Res., VOL. 118, 3357–3363, doi:10.1002/jgra.50376

Zhao, L. L., Zhang, H., and Zong, Q. G. (2017), Global ULF waves generated by a hot flow anomaly, Geophys. Res. Lett., 44, 5283–5291, doi:10.1002/2017GL073249.

**Appendix B. Supporting information**

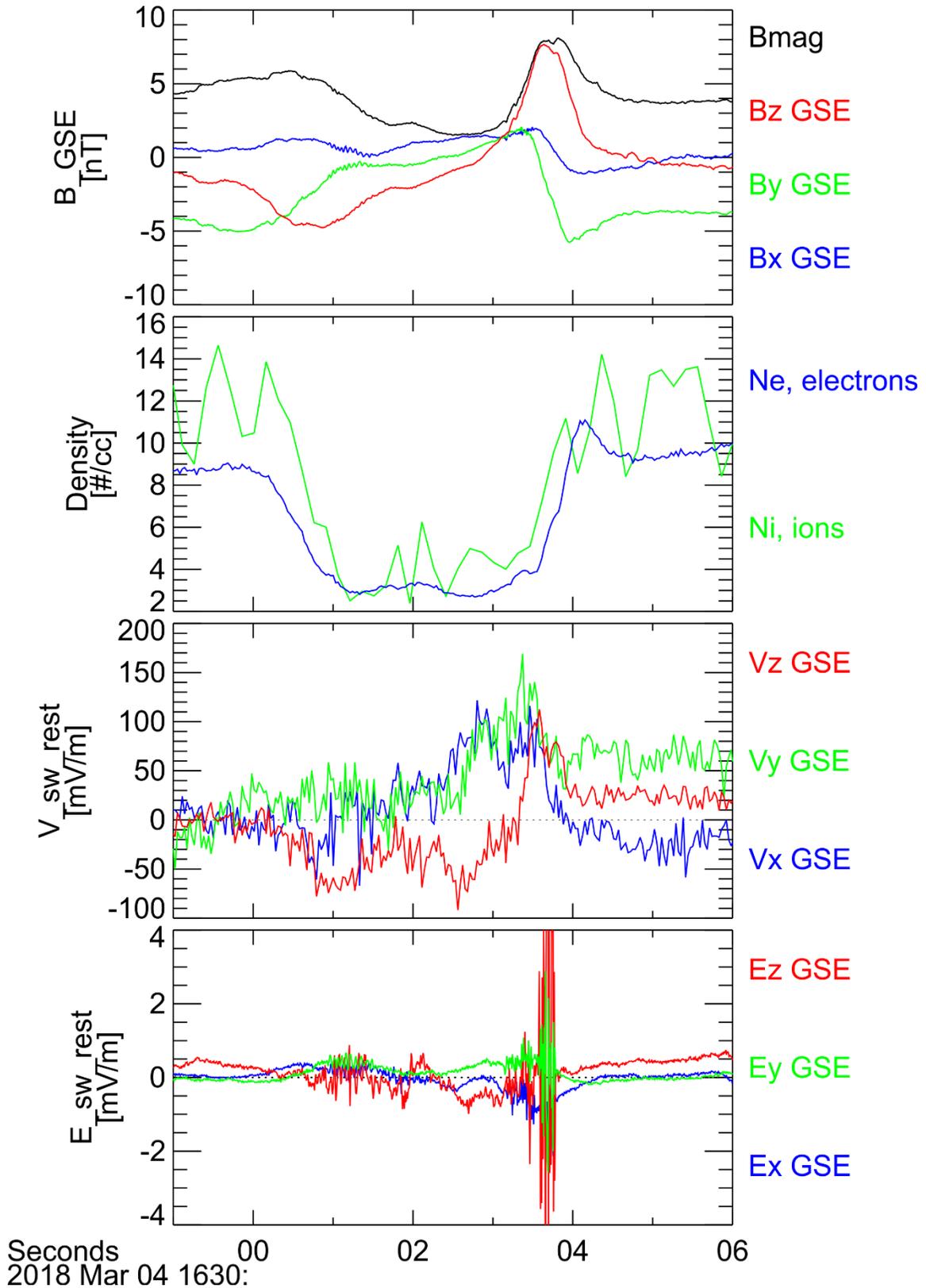

**Figure S1.** Electron bulk velocity $V_e$ in the solar wind rest frame is calculated by subtracting electron bulk velocity in the downstream background value. The electric field in the solar wind rest frame is calculated through $\mathbf{E}+\mathbf{V_{sw}}\times\mathbf{B}$.

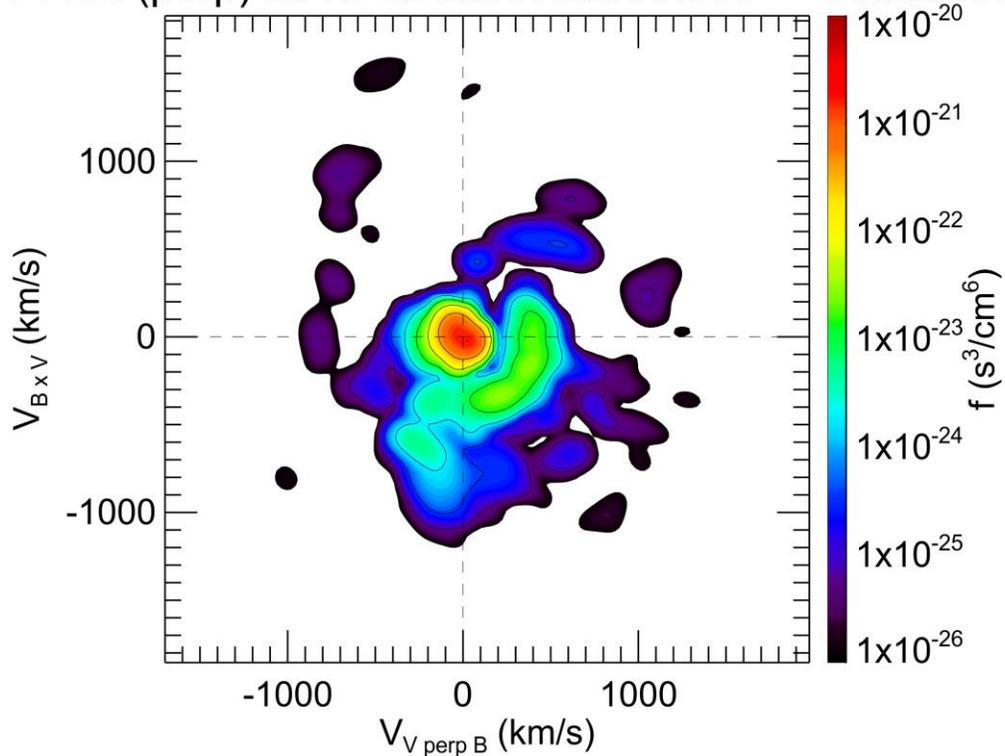

**Figure S2.** MMS1 observations of ion distribution in the perpendicular cut (the horizontal axis is the foreshock ion bulk velocity projected direction) at the time corresponding to the third vertical dotted line in Figure 5.2. The distribution was in the local solar wind rest frame, so the solar wind beam was at the origin. The foreshock ions were gyrating around the origin with a certain gyrophase corresponding to the orange curved arrow in Figure 6, i.e., reflection at the compressional boundary. In this frame, the local solar wind ions were not moving, so the convection electric field was zero and foreshock ions did not change energy during the gyration as seen from the distribution. In other reference frames, however, the shifted origin can cause energy variation during the gyration. In the solar wind rest frame, the local solar wind ions were moving sunward due to the expansion. The foreshock ions before the partial gyration or reflection (sunward) had higher energy than those after the reflection (earthward), i.e., foreshock ions lost energy through partial gyration against the convection electric field. In the spacecraft frame, the local solar wind ions were moving earthward. The foreshock ions before the reflection

(sunward) had lower energy than those after the reflection (earthward), i.e., foreshock ions gained energy through partial gyration along the convection electric field.